\documentclass[twocolumn]{aastex631}

\received{July 14, 2021}
\revised{October 12, 2021}
\accepted{March 21, 2022}
\submitjournal{The Planetary Science Journal}

\usepackage{amsmath}
\usepackage{cases}

\graphicspath{{./}{figs/}}

\shorttitle{Retention of water and carbon-rich early atmospheres}
\shortauthors{Bower et al.}

\begin{document}


\title{Retention of Water in Terrestrial Magma Oceans and Carbon-rich Early Atmospheres}

\correspondingauthor{Dan J. Bower}
\email{daniel.bower@unibe.ch}

\author[0000-0002-0673-4860]{Dan J. Bower}
\affiliation{Center for Space and Habitability, University of Bern \\
Gesellschaftsstrasse 6 \\
3012 Bern, Switzerland}

\author[0000-0003-4815-2874]{Kaustubh Hakim}
\affiliation{Center for Space and Habitability, University of Bern \\
Gesellschaftsstrasse 6 \\
3012 Bern, Switzerland}

\author[0000-0002-1462-1882]{Paolo A. Sossi}
\affiliation{Institute of Geochemistry and Petrology, Department of Earth Sciences, ETH Zurich \\
Clausiusstrasse 25 \\
8092 Zurich, Switzerland}

\author[0000-0003-3968-8482]{Patrick Sanan}
\affiliation{Institute of Geophysics, ETH Zurich \\
Sonneggstrasse 5 \\
8092 Zurich, Switzerland}



\begin{abstract}
Massive steam and CO$_2$ atmospheres have been proposed for magma ocean outgassing of Earth and terrestrial planets.  Yet formation of such atmospheres depends on volatile exchange with the molten interior, governed by volatile solubilities and redox reactions.  We determine the evolution of magma ocean--atmosphere systems for a range of oxygen fugacities, C/H ratios and hydrogen budgets that include redox reactions for hydrogen (H$_2$--H$_2$O), carbon (CO--CO$_2$), methane (CH$_4$), and solubility laws for H$_2$O and CO$_2$.  We find that small initial budgets of hydrogen, high C/H ratios, and oxidizing conditions, suppress outgassing of hydrogen until the late stage of magma ocean crystallization.  Hence early atmospheres in equilibrium with magma oceans are dominantly carbon-rich, and specifically CO-rich except at the most oxidizing conditions.  The high solubility of H$_2$O limits its outgassing to melt fractions below $\sim$30\%, the fraction at which the mantle transitions from vigorous to sluggish convection with melt percolation.  Sluggish melt percolation could enable a surface lid to form, trapping water in the interior and thereby maintaining a carbon-rich atmosphere (equilibrium crystallization).  Alternatively, efficient crystal settling could maintain a molten surface, promoting a transition to a water-rich atmosphere (fractional crystallization). However, additional processes, including melt trapping and H dissolution in crystallizing minerals, further conspire to limit the extent of H outgassing, even for fractional crystallization. Hence, much of the water delivered to planets during their accretion can be safely harbored in their interiors during the magma ocean stage, particularly at oxidizing conditions.
\end{abstract}



\section{Introduction}

The origin and distribution of light volatile elements such as C, H, and O, in the major planetary reservoirs (core, mantle, and atmosphere), regulates how terrestrial planet atmospheres form and evolve \citep[][and references therein]{GBF21}.  During planetary accretion and contemporaneous iron-core formation---both of which promote the formation of a global magma ocean---the siderophile nature of both C \citep{CWF08} and H \citep{TSH21} strongly partitions them into the core.  Nevertheless, the mantle retains some fraction of C at core-forming conditions, in addition to replenishment of both C and H after core formation \citep{H16}; hence, these volatiles subsequently participate in short- and long-term geochemical cycling \citep[e.g.,][]{DAS13}.  Here we examine how the inventories of C and H exchange between the interior and atmosphere as the magma ocean evolves after core formation.

The magma ocean outgasses volatiles as it cools from molten to mostly solid silicate rock to form a secondary atmosphere.  Therefore, chemical and physical processes operating in the magma ocean, both at its surface and at depth, exert strong controls on the size and composition of such early atmospheres.  Early atmospheres establish the initial conditions for the subsequent long-term evolution of planetary interiors and their atmospheres through interaction with the surface environment via geochemical cycles.  Therefore, magma ocean atmospheres provide the connection between the volatile budgets emplaced during planet formation and the establishment of long-term climate states.

The shock degassing of substantial H$_2$O from hydrated minerals \citep{LA82} to simulate impacts during planetary accretion motivated investigation of the blanketing effect of a steam atmosphere above the molten early Earth \citep{AM85}.  Additional justification for the widespread presence of H$_2$O and CO$_2$ is provided by experiments and accompanying equilibrium models for the devolatilization of ordinary and carbonaceous chondrites, which are shown to produce oxidized species in abundance \citep{GM77,SF10,TTS21}.  Consequently, coupled magma ocean--atmosphere models have focused almost exclusively on the role of oxidized species (H$_2$O and CO$_2$) in controlling the lifetime of magma oceans and growth of early atmospheres \citep[e.g.,][]{LMC13,HH17,SMD17,NKT19,BCB21}.

More reduced materials (such as enstatite chondrites that are thought to have contributed to Earth's accretion on isotopic grounds) may instead produce H$_2$-rich or CO-rich gases \citep{SF10}.  This has partly inspired calculations of the chemistry and cooling timescales of reducing atmospheres \citep{ZLC20,LBH21}.  However, chondrites may not be appropriate analogs for the materials that degas to produce early atmospheres around terrestrial planets \citep{SOSSI21}.  This is because mass transfer is much more efficient when materials are liquid rather than solid, allowing for the more rapid replenishment of surface material via convection and thus facilitating atmosphere formation.  This has further led coupled models to understand the astrophysical observables of magma oceans to interpret the spectroscopic data from next-generation telescopes \citep{HKA15,BLB19,BKW19,KOG20}.

The redox state (more precisely, the oxygen fugacity, $f{\rm O_2}$) of the mantle depends on pressure, temperature, and composition and dictates the speciation of outgassing products. It controls the relative fugacities of reduced and oxidized species, which is equivalent to partial pressure at the low total pressures of terrestrial atmospheres.  Core formation on Earth established $f{\rm O_2}$ around two log10 units below the iron-w\"{u}stite (IW) buffer at equilibrium at depth, based on the Fe content of the mantle and core \citep[e.g.,][]{RNM15}.  However, the volume change of the reaction between ferric and ferrous iron means that silicate melt transported to the surface defines higher $f{\rm O_2}$ than that set by core formation \citep{AFM19,HMM12}.  The surface $f{\rm O_2}$, rather than the $f{\rm O_2}$ at depth, imposes the redox state at the magma ocean--atmosphere interface and thereby controls outgassing chemistry \citep{SOSSI20}.

Evolutionary models of magma oceans are usually derived from geochemical or geodynamic considerations.  Chemical models focus on tracking the compositional evolution of cumulate mineral assemblages that form as the magma ocean cools \citep[e.g.,][]{ET08}.  However, they ignore dynamics during the crystallization process and often assume fractional crystallization.  In (stepwise) fractional crystallization, cumulates that form owing to cooling are isolated from the molten magma ocean, and the evolving composition of the melt is determined from mass balance.  By contrast, dynamic models track thermal energy transport but at the expense of reduced compositional complexity.  However, dynamic formulations that include a local representation of melt--solid separation \citep{ABE93,BSW18} can replicate fractional crystallization.  They can also replicate equilibrium crystallization, in which melt and solid freeze together before any significant relative motion has occurred.

Equilibrium chemistry calculations propose a $\sim$100 bar CO-rich atmosphere for an Earth-like magma ocean around 2200 K \citep{SOSSI20}, although it remains unclear how the atmospheric size and speciation change during the cooling and crystallization of the magma ocean.  In particular, it is unknown whether all early atmospheres are expected to be CO-rich and whether they can transition to atmospheres that are instead dominated by hydrogen species, such as H$_2$, H$_2$O, and CH$_4$---with implications for habitability and the formation of surface water oceans.  Furthermore, it remains to be determined how the transition is influenced by planetary conditions (e.g., $f{\rm O_2}$, C/H ratio, H budget), which is pertinent given the diversity of formation environments for terrestrial planets outside the solar system.

To this end, we combine mass balance and equilibrium chemistry in a self-consistent and time-evolved model to probe magma ocean--atmosphere exchange.  Unlike previous models, our model accounts for the CO$_2$/CO, H$_2$O/H$_2$, and CH$_4$/CO$_2$ ratio set by the oxygen fugacity, C/H ratio and temperature of the magma ocean, as well as their relative solubilities. We explore a range of initial endowments of the hydrogen and carbon volatile inventory and quantify the relationship between these variables and the partial pressures of the degassed species. Finally, atmospheric escape is also modeled to determine the extent to which the preferential loss of hydrogen can influence the subsequent outgassing of dissolved volatiles.

\section{Interior--Atmosphere Coupling}
\subsection{Overview}
The SPIDER code \citep{BSW18,BKW19} solves for the coupled evolution of the silicate mantle and atmosphere, where the mantle can be molten, solid, or a mixture (partial melt).  It considers interior energy transport by convection, conduction, and relative motion of melt and solid through mixing and separation and is a true 1D model in the sense that energy fluxes are determined locally \citep{ABE93,BSW18}.  Cooling of the interior is regulated by radiative transfer in the atmosphere, and the atmosphere itself grows through outgassing of the interior as the mantle solidifies owing to cooling.  We consider a two-phase system of single composition (MgSiO$_3$) for simplicity, where the thermophysical properties of solid and melt are determined from \cite{M09} and \cite{WB18}, respectively.  The melting curves adhere to peridotite melting data in the upper mantle and measurements on chondritic material in the lower mantle \citep{ABL11}; they are not perturbed according to the mantle volatile content.  The atmosphere is treated as gray with no scattering, and the two-stream approximation is applied to solve for radiative transfer \citep{AM85}; Radiation limits are not considered.  Opacities of the gas species are provided in Appendix~\ref{sect:opacity}.

We modified the SPIDER code to additionally account for mass exchange between volatiles according to equilibrium chemistry (Section~\ref{sec:chem}), as well as atmospheric escape (Section~\ref{sec:escape}).  The mathematical description of the volatile mass balance and its adherence to equilibrium chemistry is given in detail in Appendix~\ref{app:volatiles}.  For each chemical reaction, we introduce a term into the volatile mass balance to account for the exchange of mass necessary to retain chemical equilibrium between participating gas species as dictated by oxygen fugacity and temperature.  Similarly, another term in the mass balance equation tracks volatile loss due to atmospheric escape.  Hence, chemical equilibrium and escape are self-consistently determined as part of the same system of equations that are integrated to describe the coupled interior--atmosphere evolution.

Our parameter choices are similar to those in \cite{BSW18,BKW19}, so we only present pertinent details here.  We consider a planet with Earth dimensions and a magma ocean that crystallizes from the bottom up, an approach that is justified owing to the steeper melting curves compared to the mantle adiabat.  Convection and melt--solid mixing are determined by an eddy diffusivity that is based on a constant mixing length, and internal heat sources are not considered because their influence is negligible over the lifetime of a magma ocean. Viscosity controls the dynamic regime of the mantle and varies from $10^2$ Pa s for pure silicate melt to $10^{22}$ Pa s for pure solid, increasingly abruptly at a melt fraction of 40\% \citep{CCB09,BSW18}. Other controls on viscosity, such as temperature and composition, are second order and therefore neglected. The crystal size, which controls the efficiency of melt--solid separation, is set at either 1 mm or 5 cm (Appendix~\ref{sect:sep}) to simulate equilibrium or fractional crystallization, respectively.

We track the reservoir evolution of five volatile species during the magma ocean stage: H$_2$, H$_2$O, CO, CO$_2$, and CH$_4$.  For convenience, we refer to H$_2$O and H$_2$ as hydrogen species and CO, CO$_2$, and CH$_4$ as carbon species.  The dissolved content of H$_2$O and CO$_2$ in the magma ocean is determined by their solubility, and reduced species are set to zero solubility (Section~\ref{sect:sol}).  Redox reactions impose H$_2$--H$_2$O, CO$-$CO$_2$, and CO$_2$--H$_2$--CH$_4$ equilibrium (Section~\ref{sec:chem}).  Partitioning of volatiles into solids (crystals) is not considered because its effect is small compared to the partitioning of volatiles between the melt and atmosphere.  Therefore, the amount of volatiles stored in the mantle during solidification is a minimum estimate.  We simulate cooling of a magma ocean from an initial surface temperature of 2700 K until it reaches 1650 K.  For surface temperatures cooler than 1650 K the assumption of equilibrium dissolution begins to break down.  This is because a rheological transition caused by increasing crystal fraction results in mass transfer that is too sluggish to maintain equilibrium with the atmosphere; for conciseness we subsequently refer to this event as "surface lid formation."

\subsection{Solubility}
\label{sect:sol}

A solubility law relates the dissolved volatile content of a particular species in melt $X$ to its fugacity $f$.  Pressures at the magma ocean--atmosphere interface of an Earth-sized planet are expected to be around several hundred bars, which motivates our subsequent choice of solubility laws.  In the following, the dissolved content of a given volatile in the magma ocean is set solely by its fugacity at the interface and the relevant solubility law.  The volatile species \textbf{are} considered to be homogeneously distributed in the magma ocean.

\subsubsection{Solubility of H$_2$O}
\label{sect:solH2O}
We employ a general expression for solubility with a power-law dependence:
\begin{equation} 
X = \alpha f^{1/\beta},
\label{eq:sol}
\end{equation}
where $\alpha$ and $\beta$ are constants specific to a particular equilibrium between a gas species and its dissolved component.  In Equation~\ref{eq:sol}, $\alpha$ is an empirically determined coefficient that encompasses the temperature, pressure, and composition of the liquid in which the solubility is being determined, while $\beta$ reflects the stoichiometric relationship between the mole fraction of the dissolved species and the fugacity of the gas species.  Henry's law is recovered when $\beta=1$, which indicates no chemical reaction (hence no change in speciation) when the gas dissolves in the solvent.  Throughout this work we assume gas ideality such that fugacity and partial pressure are equivalent, and hence symbols for fugacity $f$ and partial pressure $p$ are used interchangeably.  In Equation~\ref{eq:sol}, the relevant partial pressure (fugacity) is determined at the interface of the magma ocean and atmosphere.

At low H$_2$O fugacities, water dissolves in silicate melts as hydroxyl (OH$^-$) group, so $\beta_{\rm H_2O}=2$ describes the solubility of H$_2$O in melt \citep[e.g.,][]{M94}.  However, this relationship breaks down at high H$_2$O fugacities because H$_2$O replaces OH$^-$ as the predominant melt species \citep[above $\sim 2000$ bars,][]{BLH02,S82}.  Therefore, a single $\beta_{\rm H_2O}$ determined by simply best-fitting data across large $f$H$_2$O ranges yields spurious values of $\beta_{\rm H_2O}$ because different reactions govern solubility at different $f$H$_2$O.

Table~\ref{tab:h2o_sol} summarizes recent experimental constraints on H$_2$O solubility for compositions, pressures, and temperatures relevant to the magma ocean--atmosphere interface of a terrestrial planet, i.e. a few hundred bars and greater than about 1500 K.  The table is ordered based on the silica content of the sample, increasing from less than 45 wt\% SiO$_{2}$ to more than 50 wt\% SiO$_{2}$.  Experiments on evolved compositions (e.g. rhyolitic/granitic melts) are not included because they are not representative of compositions during the magma ocean stage: these are instead summarized in \cite{IMT12}.

\begin{figure}[tbhp]
\plotone{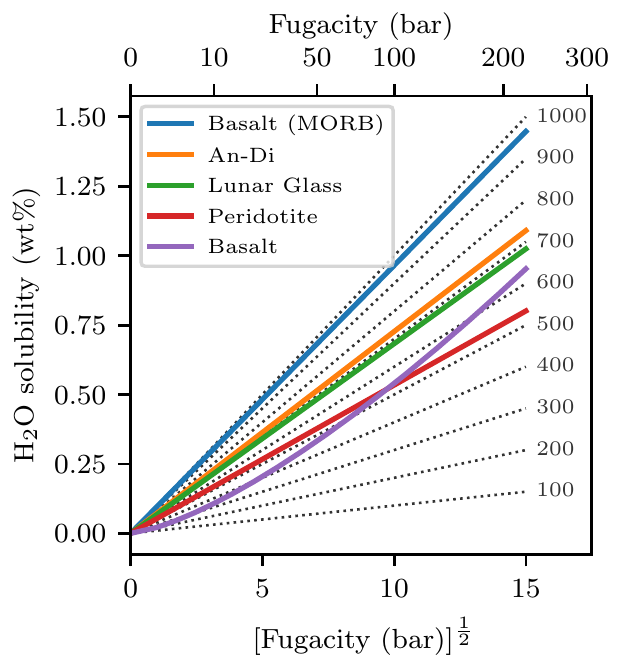}
\caption{Experimentally determined solubility data of H$_2$O for several melt compositions and temperatures, for fugacities from 0 to 200 bars (Table~\ref{tab:h2o_sol}).  Dotted lines denote $\beta=2.0$, and $\alpha$ is reported at the end of the lines (Equation~\ref{eq:sol}).}
\label{fig:sol}
\end{figure}

Peridotite is a rock composed predominantly of olivine, which is the major constituent of Earth's mantle and likely exoplanetary mantles \citep{PR19}.  The solubility of water in liquid peridotite (red line, Figure~\ref{fig:sol}) is determined at high temperature and low $f$H$_2$O, so its solubility law is most appropriate for a terrestrial magma ocean \citep{STB21}.  For the relevant pressure--temperature conditions of the magma ocean surface, the mole fraction of H$_2$O dissolved depends largely on variations in $f$H$_2$O, while the effect of temperature is a subordinate \citep{HO86}, if not poorly constrained, variable.  As such, temperature does not explicitly appear as a functional dependence in Equation~\ref{eq:sol}, and $\alpha$ is taken to be constant.
\subsubsection{Solubility of CO$_2$}
There are no experimental constraints on the solubility of CO$_2$ in peridotite, so instead we use a solubility law formulated for basalt as a proxy.  At low CO$_2$ fugacities, carbon dioxide dissolves in silicate melts as either molecular CO$_{2}$ or the CO$_3^{2-}$ group \citep{SH88}.  Because the stability of the carbonate ion increases according to melt basicity (Mg$^{2+}$, Ca$^{2+}$ cations, \cite{HB94}), the Mg-rich nature of planetary mantles ensures that it occurs exclusively as CO$_3^{2-}$ in such liquids.  Determining the mole fraction of dissolved carbonate for pressures up to 815 bars $f$CO$_2$, \cite{DSH95} constrained the solubility of CO$_2$ in mid-ocean ridge basalt (MORB) melt:
\begin{equation}
X^m_{\rm CO_2} = (3.8 \times 10^{-7}) f{\rm CO_2} \exp \left(-23\  \frac{f{\rm CO_2}-1}{10 R T} \right),
\label{eq:CO2_sol1}
\end{equation}
where $X^m_{\rm CO_2}$ is the mole fraction of CO$_2$ in the melt, $f{\rm CO_2}$ is fugacity in bars, and $T$ temperature and $R$ gas constant in SI units.  The exponential function on the right-hand side is known as the Poynting factor, which captures the influence of pressure and temperature on CO$_2$ fugacity.  The abundance of CO$_2$ (ppmw) in melt is then
\begin{equation}
X_{\rm CO_2} = 10^4 \left( \frac{4400 X^m_{\rm CO_2} }{36.594-44X^m_{\rm CO_2}} \right).
\label{eq:CO2_sol2}
\end{equation}
This solubility law compares favorably to experimental results at 10 kbar \citep{PHH91}.  At water concentrations less than about 3 wt\% there is a negligible influence of the presence of dissolved H$_2$O on CO$_2$ solubility \citep{DSH95}.  Above about 6 wt\%, water can enhance the solubility of CO$_2$ by around a factor of two \citep{IMT12}, although CO$_2$ remains insoluble compared to H$_2$O.  In our model, the water concentration only increases beyond 3 wt\% at the end of the magma ocean stage for the most oxidizing condition and largest initial hydrogen budget; otherwise, the water concentration is significantly less.  Hence, we ignore the minor influence of both carbon and hydrogen on each other's solubility.

\subsubsection{Solubility of H$_2$, CO, and CH$_4$}
We assume that the solubilities of H$_2$, CO, and CH$_4$ are negligible compared to H$_2$O and CO$_2$, respectively, and therefore set $\alpha_{\rm H_2}=\alpha_{\rm CO}=\alpha_{\rm CH_4}=0$ (Equation~\ref{eq:sol}).  This is motivated by high-pressure experiments that find that the equilibrium constants for the dissolution of H$_2$ gas in basaltic melts result in solubilities more than two orders of magnitude lower than that for H$_2$O \citep{LIDT15,HWA12}. Similarly, the mole fraction of reduced carbon dissolved in silicate melts is typically one to two orders of magnitude lower than for CO$_2$ \citep{YNN19}.

Methane dissolved in silicate melts has been identified by spectroscopy in quenched glasses \citep{MKCF11, AHW13, WRJ13, AHS15, DHJ19}. These studies together demonstrate that the fraction of dissolved methane increases in melts equilibrated under increasingly reducing, high-pressure and H-rich conditions.  To quantify the solubility of methane in the melt at a given methane fugacity, \cite{AHW13} performed experiments with Fe-free basaltic melts at pressures between 0.7 and 3 GPa.  The equilibrium constant of the reaction CH$_{4}$(l) = CH$_{4}$(g) results in an order of magnitude less dissolved C for a given $f$CH$_{4}$ compared to an equivalent $f$CO$_{2}$. Because estimated CH$_{4}$ contents are lower than several hundred ppmw even at $f$CH$_{4}$ of 1000 bars \citep{AHW13}, its solubility is neglected. It is noteworthy that solubility data for CH$_{4}$ in silicate melts at low total pressures relevant to magma oceans ($<$0.7 GPa) are lacking, preventing an accurate assessment of its solubility behavior.

\begin{deluxetable*}{LLlLLll}
\centerwidetable
\tablecaption{H$_2$O Solubilities Constrained by Experimental Studies\label{tab:h2o_sol}}
\tablewidth{0pt}
\tablehead{
\colhead{${\alpha}^\ddagger$} & \colhead{$\beta$} & \colhead{Composition} & \colhead{Temp. (K)} & \colhead{Pres. (bars)} & \colhead{$f{\rm O_2}$ ($\Delta$IW)} & \colhead{Reference}}
\decimals
\startdata
534 & 2.0 & Peridotite & 2173 & 1.013 & --1.4 to +6.5 & \cite{STB21}\\
683 & 2.0 & Lunar Glass & 1623 & 1.013 & --3.0 to +4.8 & \cite{NBB17}\\
727 & 2.0 & Anorthite-Diopside & 1623 & 1.013 & --3.0 to +4.8 & \cite{NBB17}\\
965 & 2.0 & Basalt (MORB) & 1473 & 176-717 & +3.9 to +13.2 & \cite{DSH95}\\
1007 & 2.0 & Basalt (MORB) & 1473 & 503-2021 & +3.5 and +7.9 & \cite{BLH02}\\
215 & 1/0.7 & Basalt & 1373 & 1034-6067 & unbuffered & \cite{WH81}$^\dagger$
\enddata
\tablecomments{$^\dagger$ Used in several modeling studies, e.g. \cite{LMC13}, \cite{SMD17}, \cite{NKT19}, and \cite{BKW19}.  $^\ddagger$ $\alpha$ has units of ppmw/bar$^{1/\beta}$ and where relevant is determined by refits to existing data by constraining $\beta=2$ and no solubility at zero fugacity.  We only consider experimental studies of basaltic or more mafic melts (i.e., Mg and Fe rich) at relatively low $f{\rm H_2O}$ ($\lesssim$2000 bars).} 
\end{deluxetable*}

\subsection{Redox Reactions}
\label{sec:chem}
When at equilibrium, the magma ocean and atmosphere must be considered as a single thermodynamic unit.  This is because, based on the scaling of convective velocity \citep{VSS00}, the magma ocean mixing timescale is a few weeks, which is considerably shorter than the cooling timescale of around 1 Myr.  Accordingly, $f{\rm O_2}$ defined by the magma ocean at its surface is equivalent to the oxygen partial pressure (or fugacity) in the atmosphere. Therefore, the fugacity ratio of gaseous redox couples (H$_2$/H$_2$O and CO/CO$_2$), as well as CO$_2$--H$_2$--CH$_4$, satisfy chemical equilibrium at the $T$ and $f{\rm O_2}$  imposed by the magma ocean at its surface.  We consider three reactions, where all species exist in the gas phase, albeit partially dissolved in the melt according to their respective solubilities (Sect.~\ref{sect:sol}).  For the hydrogen redox couple,

\begin{equation}
\begin{aligned}
\rm{H}_2 +  0.5 \rm{O}_2  &= \rm{H}_2O,\\
\log_{10}{\rm K_{eq}} &= \frac{13152}{T} - 3.039,
\end{aligned}
\label{eq:h_react}
\end{equation}
for the carbon redox couple,

\begin{equation}
\begin{aligned}
\rm{CO} +  0.5 \rm{O}_2  &= \rm{CO}_2,\\
\log_{10} {\rm K_{eq}} &= \frac{14468}{T} - 4.348,
\end{aligned}
\label{eq:c_react}
\end{equation}
and for CO$_2$--H$_2$--CH$_4$,

\begin{equation}
\begin{aligned}
\rm{CO}_2 + 2\rm{H}_2 &= \rm{CH}_4 + \rm{O}_2,\\
\log_{10} {\rm K_{eq}} &= \frac{-16276}{T} - 5.4738.
\end{aligned}
\label{eq:ch4_react}
\end{equation}

The equilibrium constants $\rm{K_{eq}}$ for the redox couples are determined from the Gibbs free energy of reaction using data from the JANAF database fit for temperatures between 1500 and 3000 K \citep{JANAF}.  The equilibrium constant for the CO$_2$--H$_2$--CH$_4$ reaction is taken from the IVTANTHERMO database fit for temperatures between 300 and 2000 K \citep{SF17}.  The most abundant redox-sensitive element in Earth's mantle is iron, so the amount of oxygen that is free to participate in reactions will be approximately regulated by Gibbs free energy changes along the IW buffer during the magma ocean stage.  Due to the presence of melt and the difficulty of determining the thermodynamic properties of a nonstoichiometric phase (w\"{u}stite), it is preferable to consider equilibrium of solid metallic Fe with liquid FeO (the IW buffer), where the Gibbs free energy of reaction constrains $f{\rm{O_2}}$ \citep{OE02}:

\begin{equation}
\begin{aligned}
&\rm{Fe} +  0.5 \rm{O}_2  = \rm{FeO},\\
&\log_{10} {f{\rm O_2, IW}} = \frac{-244118+115.559 T-8.474 T \ln T}{0.5 \ln(10) R T}.
\end{aligned}
\label{eq:fO2}
\end{equation}
It is convenient to define oxygen fugacity in log10 units relative to the IW buffer ($\Delta {\rm IW}$):

\begin{equation}
    \Delta{\rm IW} = \log_{10} f{\rm O_2} - \log_{10} f{\rm O_2, IW}.
\end{equation}
In our nominal models we set $\Delta {\rm IW}=0.5$ based on a recent determination of the inferred oxygen fugacity of Earth's magma ocean at its surface using the Fe$^{3+}$/Fe$^{2+}$ ratio from a global compilation of peridotites \citep{SOSSI20}.  Hence, by fixing $f{\rm O_2}$ relative to the IW buffer, H$_2$O/H$_2$ (Equation~\ref{eq:h_react}) and CO$_2$/CO (Equation~\ref{eq:c_react}) are constrained as a function of temperature, and by extension CH$_4$/CO$_2$ (Equation~\ref{eq:ch4_react}).  Therefore, while solubility controls the abundance of species in the atmosphere versus the interior, redox reactions dictate the relative abundance of oxidized to reduced species in the atmosphere.  The thermodynamic coupling between the atmosphere and the interior is set by the surface temperature at the magma ocean--atmosphere interface.  Appendices~\ref{app:volmass} and \ref{app:chemreact} describe how reactions are self-consistently incorporated into the time stepper that evolves the magma ocean--atmosphere system.

\subsection{Atmospheric Escape}
\label{sec:escape}
Atmospheric escape from terrestrial planets is challenging to parameterize given the complex interaction of thermal and nonthermal processes, as well as uncertainty regarding the history and details of the stellar environment and planetary magnetic field \citep[e.g.,][]{GAB20}.  In the context of this study, we concern ourselves with the impact of H$_2$ loss from the atmosphere on the evolution of the magma ocean--atmosphere system.  We focus on hydrogen since it is the lightest atmospheric component and therefore the most prone to escape. Escape of H$_2$ may be buffered by outgassing from the interior, resulting in efficient depletion of hydrogen from the magma ocean.  We apply a model that obeys an upper limit for H$_2$ loss based on the diffusion rate of H$_2$ through a hydrostatic carbon-dominated atmosphere \citep{ZGC19}.  Below this limit, the escape flux follows the energy-limit that considers the stellar energy input necessary for a volatile to escape Earth's gravitational field.  Therefore, the H$_2$ escape flux is \citep{ZGC19}

\begin{equation}
\phi_{\rm H_2} \approx \frac{\gamma 10^{16} \rm{[VMR]}_{\rm H_2} S}{N_A \sqrt{1+0.006S^2}}\qquad \text{moles}\ {\rm m}^{-2}\ {\rm s}^{-1},
\label{eq:escape}
\end{equation}
where $N_A$ is the Avogadro constant and ${\rm [VMR]}_{\rm H_2}$ is the volume mixing ratio of H$_2$ that provides coupling between the evolving atmospheric speciation and H$_2$ escape.  Here $\gamma$ is a scaling factor (unity by default), and $S$ is the normalized solar irradiation according to Earth's present-day value:
\begin{equation}
S(t) = F_{\rm XUV} / F_{\rm XUV \odot},
\end{equation}
where XUV combines the influence of X-ray, extreme ultraviolet, and far ultraviolet radiation.  We adopt an upper estimate of $S=40$ at 4.0 Ga \citep{TJG15}.  For ${\rm [VMR]}_{\rm H_2}$ of 30\% and $\gamma=1$, the H$_2$ mass-loss rate is 2.1$\times10^5$ kg s$^{-1}$ (Equation~\ref{eq:masslossrate}).  Both the escape prefactor $\gamma$ (constant for a given case) and ${\rm [VMR]}_{\rm H_2}$ (time dependent) scale the mass-loss rate linearly.

\section{Results}
\subsection{Outgassing of Hydrogen (One Earth Ocean)}
\label{sect:ref}
\begin{figure*}[tbhp]
\includegraphics[width=\textwidth]{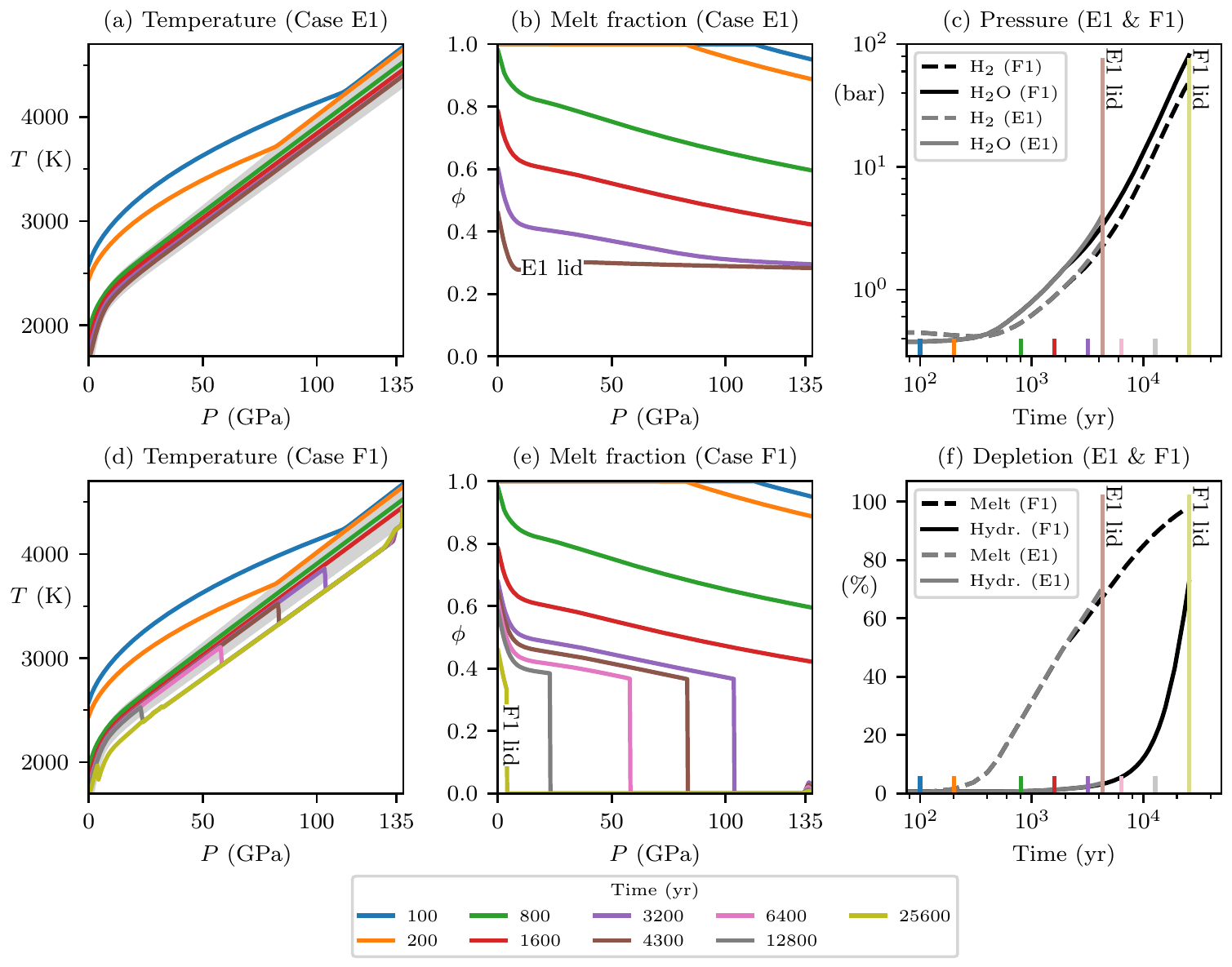}
\caption{Coupled interior and atmosphere evolution during magma ocean outgassing at $\Delta$IW=0.5 with a hydrogen budget of one Earth ocean (no carbon), where E1 shows equilibrium crystallization and F1 shows fractional crystallization.  For E1 and F1: (a, d) Interior temperature with gray showing the mixed phase region where melt and solid coexist; (b, e) interior melt fraction.  (c) Atmospheric partial pressure of H$_2$ and H$_2$O for E1 and F1, where the pressures of F1 (black lines) initially follow E1 (gray lines) before departing at late time.  Colored vertical ticks and lines show times in panels (a), (b), (d), and (e), and labels show when a surface lid forms for E1 and F1.  (f) Interior depletion of melt (`Melt') and hydrogen (`Hydr.') for E1 and F1.}
\label{fig:ref}
\end{figure*}

We first present magma ocean outgassing models (Cases E1 and F1) where the total hydrogen (H) budget is equivalent to the mass of H in Earth's present-day water ocean (hereinafter one ocean) (Figure~\ref{fig:ref}).  This budget is equivalent to a fully outgassed atmosphere of 270 bars for a single species atmosphere consisting of H$_2$O.  These models allow us to describe H$_2$ and H$_2$O outgassing before subsequently introducing further complexity in the form of carbon species, redox variation, and atmospheric escape.  The total budget of H is partitioned between a reduced (H$_2$) and an oxidized (H$_2$O) species according to the $f{\rm O_2}$ buffer set at $\Delta$IW=0.5, with an initial surface temperature of 2700 K.  We recall that in our model H$_2$ has zero solubility in the melt and so can only reside in the atmosphere.  The only parameter difference between Case E1 (hereinafter E1) and Case F1 (hereinafter F1) is the crystal size, where $a=$ 1 mm for E1 and $a=$ 5 cm for F1.  Both of these crystal sizes are reasonable for a crystallizing magma ocean and demonstrate the end-member scenarios of equilibrium and fractional crystallization that arise depending on the efficiency of melt--solid separation, which scales according to the crystal size squared (Appendix~\ref{sect:sep}).

Both E1 and F1 chart a similar thermal evolution when the melt fraction is high.  Cooling proceeds quickly since most H remains in the magma ocean in the form of H$_2$O; this keeps the atmospheric pressure low ($<$ 1 bar).  The atmospheric composition (mole fraction of H$_2$ compared to H$_2$O) is determined by chemical equilibrium, where $f{\rm O_2}$ decreases with decreasing surface temperature (Equation~\ref{eq:h_react}) but parallels the IW buffer.  Vigorous convection keeps crystals mostly in suspension assuming efficient re-entrainment of crystals \citep{SOS93,SS93} and the interior is approximately adiabatic.  Due to the curvature of the melting curves and the two-phase adiabat, the deepest mantle has the lowest melt fraction and hence reaches the `rheological transition' first.  The rheological transition is defined by an abrupt increase in viscosity around 30--40\% melt fraction, where the formation of an interconnected solid matrix begins to dictate the convective timescale.  The "rheological front" is the interface between melt-dominated dynamics above (lower pressure) and solid creep below (higher pressure), and it moves upward through the mantle as cooling proceeds.  Hence, convection in the deep mantle becomes sluggish, akin to the present-day solid mantle, acting as a brake on deep mantle cooling \citep{AML16}.  Gravitational separation of melt and solid becomes a dominant driver of energy transport at the rheological transition \citep{ABE93}.  Hence, the E1 and F1 models exhibit different interior and outgassing evolution because of the timescale of melt--solid separation (percolation) compared to net cooling rate.

E1 represents equilibrium crystallization, where the timescale for melt--solid separation is longer than the timescale for the advancement of the rheological front \citep{TM90,SS293}.  The rheological front rapidly advances through the mantle, causing the thermal profile to collapse on top of the rheological transition (Figure~\ref{fig:ref}b).  When it reaches the surface, it initiates lid formation since convection has become so sluggish that heat transport by convection to the surface cannot prevent top-down cooling.  Lid formation brings about the end of interior--atmosphere dissolution equilibrium (`E1 lid' at 4.3 kyr; Figure~\ref{fig:ref}).  At this time, the mantle is around 30\% molten since it remains pinned at the rheological transition because it has not had time to undergo any significant differentiation through melt--solid separation prior to the formation of a surface lid.  Therefore, the mantle contains sufficient melt to trap the majority of H as dissolved H$_2$O beneath the surface lid (Figure~\ref{fig:ref}f).  Hence, the subsequent interaction of these sequestered volatiles with the surface and atmosphere will be regulated by geological processes operating over long timescales (millions to billions of years), rather than by dissolution equilibrium with a comparatively short-lived magma ocean.

By contrast, for F1 the melt percolation velocity keeps pace with the upward velocity of the rheological front.  This enables the complete solidification of deep mantle layers below the rheological front by crystal settling, as well as stalling the upward progression of the rheological front.  This occurs because efficient upward draining of melt keeps the upper regions molten at the expense of cooling and fully crystallizing the lowermost mantle (Figure~\ref{fig:ref}e).  Therefore, dissolution equilibrium with the atmosphere is maintained while crystals form and are displaced to depth.  Hence, F1 represents fractional crystallization, where formed solids deep in the mantle can be considered chemically isolated from the molten reservoir above.  Below the rheological transition, efficient melt--solid separation causes the thermal profile to abruptly drop to the solidus, where now subsolidus cooling is greatly restricted by the viscous transport timescale and partly by the core that buffers the cooling of the mantle (Figure~\ref{fig:ref}d).  This ultimately enables a substantial reduction of the melt reservoir ($98\%$ crystallized) at a surface temperature of 1650 K compared to E1, where the crystal fraction only reaches 70\% (Figure~\ref{fig:ref}f).  Due to the high solubility of H$_2$O in melt, for F1 the additional $\approx 30\%$ crystallization compared to E1 further depletes H from the interior by $65\%$ before a surface lid forms.  We note that fractional crystallization could also occur for E1 after the surface lid has formed, but in this case the melt reservoir is separated from the atmosphere by a surface lid, and therefore outgassing is not driven by dissolution equilibrium.

The atmospheric pressures of H$_2$O and H$_2$ increase in unison during cooling (Figure~\ref{fig:ref}c) because the relative abundance of these species is controlled by chemical equilibrium at $\Delta$IW=0.5 (Equation~\ref{eq:h_react}).  Since H$_2$O is highly soluble in silicate liquid, early in the lifetime of the magma ocean (prior to 1 kyr while melt fractions are high) the atmospheric mass remains mostly constant and the partial pressures are controlled by the temperature dependence of the $f{\rm O_2}$ buffer and equilibrium constant (Figure~\ref{fig:ref}c, Equation~\ref{eq:h_react}).  After 1 kyr, the extent to which H outgasses is proportional to the degree of crystallinity (i.e., continued cooling and crystallization) according to the lever rule.  Despite the mantle reaching 70\% crystal fraction for E1 before a surface lid forms, the atmospheric pressures of H$_2$O and H$_2$ only reach 4 and 2.4 bars, respectively (E1 lid, Figure~\ref{fig:ref}c).  For F1, the atmospheric pressures of H$_2$ and H$_2$O (and hence their depletion in the magma ocean) track E1 until around 2 kyr (gray lines, Figure~\ref{fig:ref}c,f).  At this time, efficient melt--solid separation enables growth of a more substantial H$_2$O (80 bars) and H$_2$ (50 bars) atmosphere prior to surface lid formation ("F1 lid", Figure~\ref{fig:ref}c).  It takes around 4.3 and 26 kyr for a surface lid to form for E1 and F1, respectively, which compares favorably to a timescale of 10 kyr determined by a model with a more advanced treatment of radiative transfer \citep{LBH21}.  When a surface lid forms, the atmospheric pressure of E1 and F1 differs by more than an order of magnitude.

\subsection{Outgassing of Hydrogen (1--10 Earth Oceans)}
\begin{figure}[tbhp]
\plotone{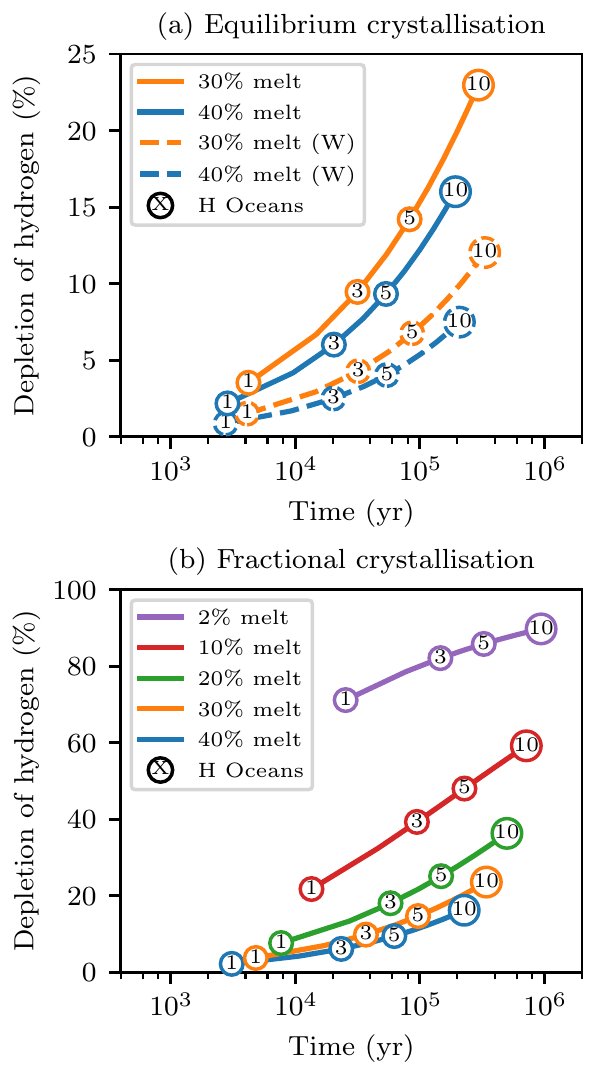}
\caption{Depletion of hydrogen (H) from the interior during magma ocean outgassing at $\Delta$IW=0.5 for H budgets from 1 to 10 Earth oceans: (a) equilibrium crystallization, and (b) fractional crystallization.  Lines show constant melt fraction, and circles show the H budget.  Magma ocean stage ends at around 30\% melt for equilibrium crystallization and 2\% melt for fractional crystallization.  Cases E1 and F1 (Figure~\ref{fig:ref}) correspond to 1 H ocean for equilibrium and fractional crystallization, respectively.  Dashed lines in panel (a) with a `W' suffix denote models that only include H as H$_2$O, i.e. a completely oxidized mantle.}
\label{fig:compare_depletion}
\end{figure}

Planet formation models predict substantial delivery of water to the inner solar system \citep{RI17,RQL07b}, which can account for the 1--10 oceans of water currently stored in Earth \citep{LGR98}.  Recent geochemical estimates propose that around 2.5 oceans are currently stored in the mantle \citep{MARTY12,H18}.  Therefore, we now present cases the same as E1 and F1 but with an initial H budget up to 10 Earth oceans.  Since the main features of the coupled evolution of the interior--atmosphere are described for E1 and F1 (Section~\ref{sect:ref}), only differences relative to these fiducial cases are now presented.  Figure~\ref{fig:compare_depletion} reveals the interior depletion of H as a percentage of the H inventory, where the circles with a solid line and containing a 1 are E1 (Figure~\ref{fig:compare_depletion}a) and F1 (Figure~\ref{fig:compare_depletion}b).

For both equilibrium and fractional crystallization, as the H inventory is increased from 1 to 10 oceans, the magma ocean lifetime increases by around two orders of magnitude, reaching a maximum duration slightly less than 1 Myr.  This is because the optical depth of the atmosphere scales with pressure, and a larger inventory results in a larger outgassed atmosphere and hence a higher atmospheric pressure.  Therefore, due to the thicker atmosphere, the magma ocean takes longer to cool before the rheological front reaches the surface and initiates lid formation.  In addition, the interior depletion of H increases for larger total inventories as a direct consequence of power-law solubility where the exponent $\beta=2$.  This is because, during cooling, the atmospheric reservoir of H grows with the water mole fraction to the second power (see Appendix~\ref{app:simple} for analysis).  Hence, for H outgassing, both the pressure of the atmosphere and relative depletion of the interior increase with the total inventory at a given melt fraction.  For a volatile species that obeys a linear solubility relation ($\beta=1$), its relative depletion at a given melt fraction is independent of its total inventory (Appendix~\ref{app:simple}). Water is expected to transition to such a linear solubility above around 10 ocean masses on Earth owing to the prevalence of dissolved H$_2$O.

For equilibrium crystallization, lid formation occurs soon after 30\% melt is reached (Figure~\ref{fig:compare_depletion}a).  For all equilibrium cases, melt--solid separation is slow compared to the cooling timescale, so when a lid forms, the melt fraction is only marginally reduced for cases with a long compared to a short cooling timescale.  At 30\% melt, a maximum of around 23\% of the total H budget is outgassed for 10 oceans, although this decreases to about 4\% for 1 ocean (Figure~\ref{fig:compare_depletion}a).  For fractional crystallization, the melt fraction decreases below 30\% until lid formation occurs around 2\% melt.  Compared to equilibrium crystallization, this extra depletion of melt drives outgassing of hydrogen to around 80\% of its total budget (Figure~\ref{fig:compare_depletion}b).  Regardless of total inventory, depletion must tend to 100\% when the melt fraction reaches zero because we assume that no volatiles are retained in solids.  This explains the flattening of the depletion curve for 2\% melt fraction compared to, for example, 10\% melt fraction.

The magma ocean lifetimes for $\Delta{\rm IW}=0.5$ cases (solid lines, Figure~\ref{fig:compare_depletion}a) are comparable to cases that are fully oxidized where H can only exist as water (W cases, dashed lines, Figure~\ref{fig:compare_depletion}a).  This is because for all cases H$_2$O dominates the opacity of the atmosphere.  However, H depletion of the $\Delta$IW=0.5 cases and that of the fully-oxidized (W) cases differ by up to about 11\%.  This is because at $\Delta$IW=0.5 outgassing of H$_2$O must be accompanied by commensurate growth of H$_2$ in the atmosphere to maintain chemical equilibrium.  Since the H budget is conserved, increasing H$_2$ in the atmosphere necessitates a decrease in the H$_2$O reservoir dissolved in the magma ocean.  In short, H$_2$O solubility determines H$_2$O outgassing, but oxygen fugacity dictates the additional depletion of interior hydrogen to maintain equilibrium between H$_2$ and H$_2$O in the atmosphere.  

\subsection{Outgassing of Hydrogen and Carbon}
\begin{figure*}[tbhp]
\centering
\includegraphics[height=0.9\textheight]{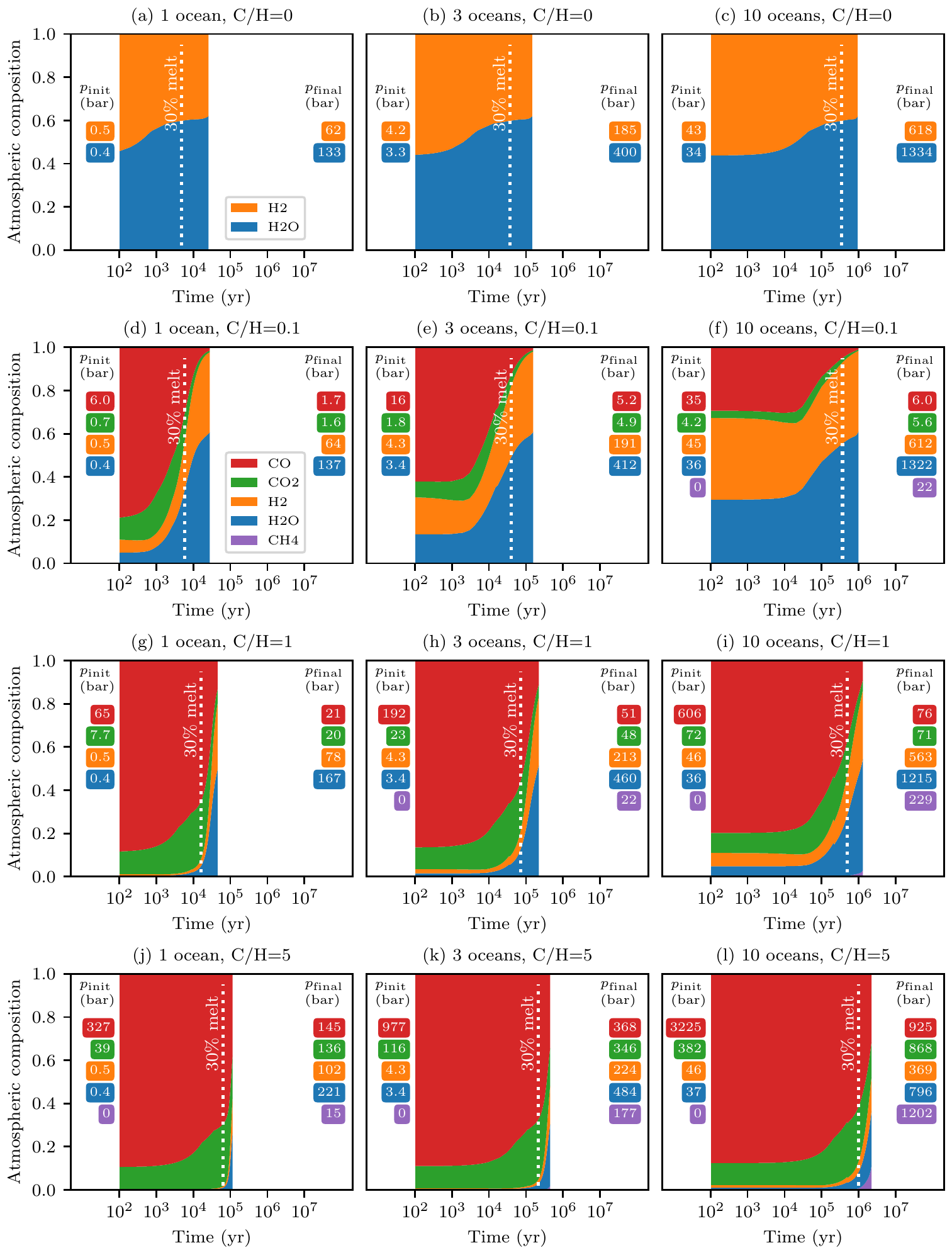}
\caption{Atmospheric composition during magma ocean outgassing at $\Delta$IW=0.5.  Columns from left to right show hydrogen (H) budgets of 1, 3, and 10 oceans, respectively.  Rows from top to bottom show C/H by weight of 0, 0.1, 1, and 5.  Magma ocean lifetime corresponds to the termination of the colored region (at $\approx$ 2\% melt for fractional crystallization, between $10^4$ and $10^7$ yr).  White dotted lines correspond to 30\% melt, which indicates both the duration and atmospheric composition if instead the mantle underwent equilibrium crystallization.  Initial partial pressures of volatile species ($p_{\rm init}$) are for a completely molten magma ocean and $T_{\rm surf}=2700$ K.  Final partial pressures ($p_{\rm final}$) are for complete solidification and outgassing of all volatile species and $T_{\rm surf}=1400$ K.}
\label{fig:ch_ratio}
\end{figure*}

In contrast to the high solubility of water in silicate melt, carbon dioxide is relatively insoluble \citep[e.g.,][]{DSH95}.  Hence, due to the high abundance of C in rocky planets, the presence of gaseous CO, CO$_2$, and CH$_4$ may regulate the lifetime of a magma ocean and thereby the timescale over which H outgasses.  Carbon also suppresses the outgassing of H through its influence on the mean molar mass of the atmosphere \citep{BKW19}.  Therefore, we now present fractional crystallization cases that additionally include C species (CO$_2$, CO, and CH$_4$), with their relative abundances in the gas phase determined by redox reactions.  The initial carbon-to-hydrogen ratio (C/H by weight) is varied from 0.1 to 5, which is a range compatible with C/H ratios in chondritic meteorites and in the bulk silicate Earth (1.1 and 1.4 by mass, respectively \citep{H16}).

We independently verified the results of our model by tabulating the final total pressure and composition (in terms of moles of H, C, and O) of select outgassed atmospheres calculated using our model (Appendix~\ref{app:factsage}). These were used as input parameters for the Equilib module of FactSage 8.0 \citep{bale2009}, which calculates the equilibrium partial pressures of gases and stable condensed phases using a Gibbs free energy minimizer with a database of more than 40 gas species in the H-C-O system. For the sake of comparison, we assumed ideal gas behavior. Provided that graphite is not predicted to precipitate (Section~\ref{sect:precipitation}) the agreement with our models, which utilize a comparatively simple chemical network and five gas species, is commendable; the partial pressures differ by at most a few percent (Table~\ref{table:factsage}).

\begin{figure*}[tbhp]
\centering
\includegraphics[height=0.95\textheight]{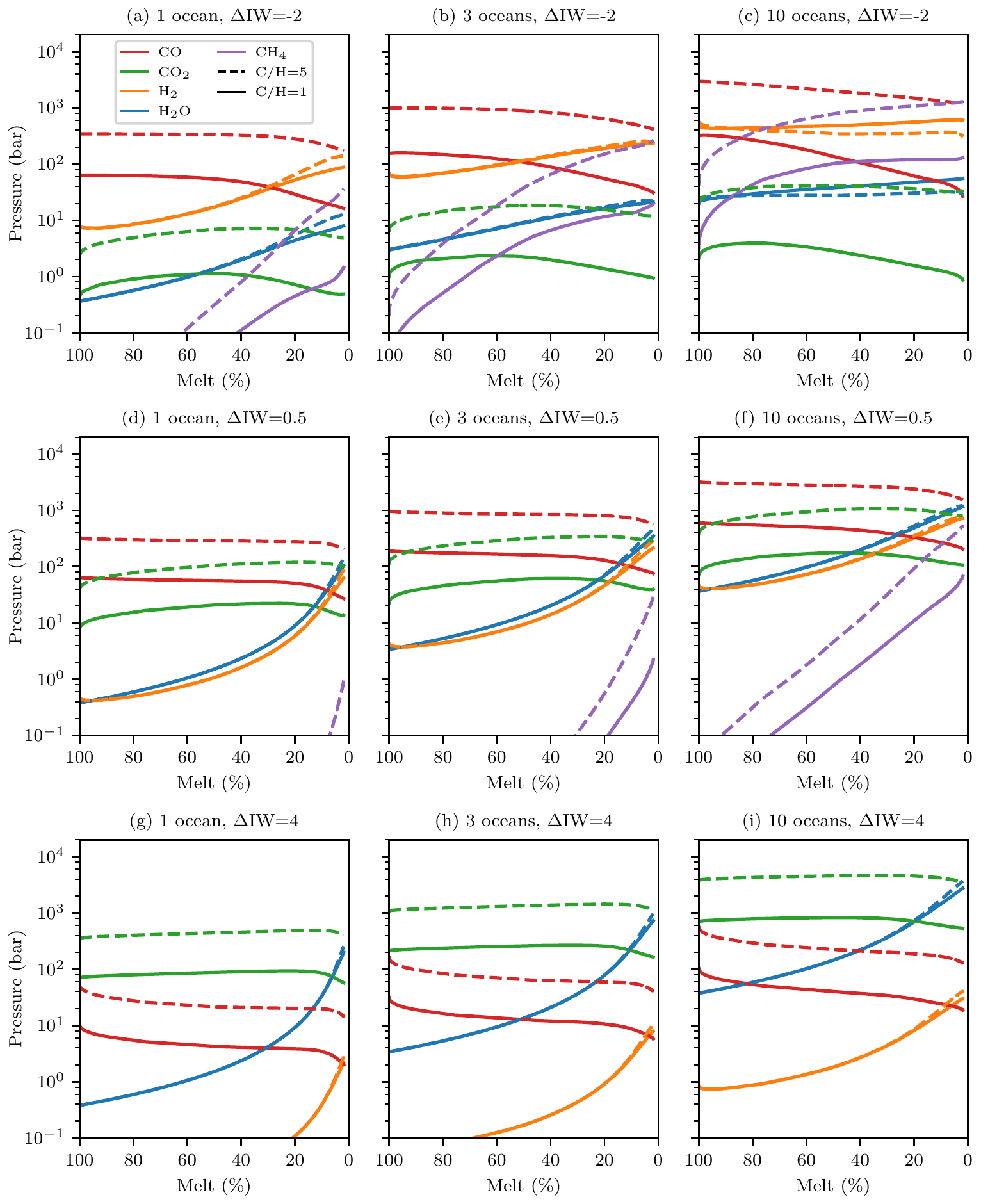}
\caption{Atmospheric pressures during magma ocean outgassing for different C/H and $f{\rm O_2}$.  Columns from left to right show hydrogen (H) budgets of 1, 3, and 10 oceans, respectively.  Rows from top to bottom show $f{\rm O_2}$ of $\Delta$IW=$-2$, $\Delta$IW=0.5, and $\Delta$IW=4.  Colors denote volatiles (red: CO; green: CO$_2$; orange: H$_2$; blue: H$_2$O; purple: CH$_4$, where solid lines show C/H=1 and dashed lines show C/H=5.}
\label{fig:ch_ratio_partial}
\end{figure*}

Figure~\ref{fig:ch_ratio} compares the influence of the C/H ratio on magma ocean outgassing for an H budget of 1, 3, and 10 oceans.  Cases F1 (Figure~\ref{fig:ch_ratio}a), F3 (Figure~\ref{fig:ch_ratio}b), and F10 (Figure~\ref{fig:ch_ratio}c) serve as carbon-free reference cases.  We first focus attention on increasing C/H for an H budget of one ocean (left column, Figure~\ref{fig:ch_ratio}).  Even for an addition of only 10\% C by weight (i.e., C/H=0.1), the atmosphere is dominated by C species (almost 90\% volume mixing ratio) for a fully molten mantle.  This is due to the low solubility of CO$_2$ compared to H$_2$O, coupled with equilibrium chemistry that establishes the partial pressure of CO around a factor of seven greater than the partial pressure of CO$_2$.  During magma ocean cooling, the CO pressure in the atmosphere steadily decreases.  This occurs as a result of the compounding effects of the equilibrium constant of Equation~\ref{eq:c_react}, which results in a decrease of $f{\rm CO}/f{\rm CO_2}$ as the surface temperature decreases, and the continued outgassing of CO$_2$ as the melt fraction decreases.  The latter also drives an initial increase in CO$_2$ pressure, although the CO$_2$ pressure decreases later once H begins to outgas in earnest.  Hence, even though the ever-decreasing melt fraction drives outgassing of both H and C throughout the magma ocean evolution, the partial pressure of volatile species can either mostly decrease (CO), mostly increase (H$_2$, H$_2$O), or increase and then decrease (CO$_2$) (Figure~\ref{fig:ch_ratio_partial}).

For an H budget of one ocean, the initial pressure of H$_2$ and H$_2$O is 0.5 and 0.4 bars, respectively, regardless of C/H.  However, at higher initial C/H ratios, the H species constitute a smaller fraction (volume mixing ratio) of the atmosphere owing to the relative insolubility of C-bearing species compared to H$_2$O.  For C/H=5, hydrogen species constitute less than 1\% of the atmosphere during the early evolution of the magma ocean ($>$30\% melt).  For this case, the mixing ratio of H species increases by a factor of 230 as the magma ocean cools from fully molten to fully solid (Figure~\ref{fig:ch_ratio}j).  In contrast, for C/H=0.1 the mixing ratio of H species only increases by a factor of about nine (Figure~\ref{fig:ch_ratio}d).  Even though the total inventory of H is fixed, the final outgassed pressures of H$_2$ and H$_2$O depend on C/H, increasing from 62 and 133 bars (C/H=0) to 102 and 221 bars (C/H=5.0), respectively.  White dotted lines in Figure~\ref{fig:ch_ratio} denote a melt fraction of 30\%, which corresponds to when a surface lid forms for equilibrium crystallization.  Hence, for equilibrium crystallization, outgassed atmospheres are usually more dominated by C species versus H species.  For C/H=0.1, the mixing ratio of H species increases by about a factor of two from 30\% melt to 2\% melt (Figure~\ref{fig:ch_ratio}d).  In contrast, for C/H=5, H species increase from about 1\% to more than 50\% over the same range of melt fraction (Figure~\ref{fig:ch_ratio}j).

\begin{figure}[tbhp]
\plotone{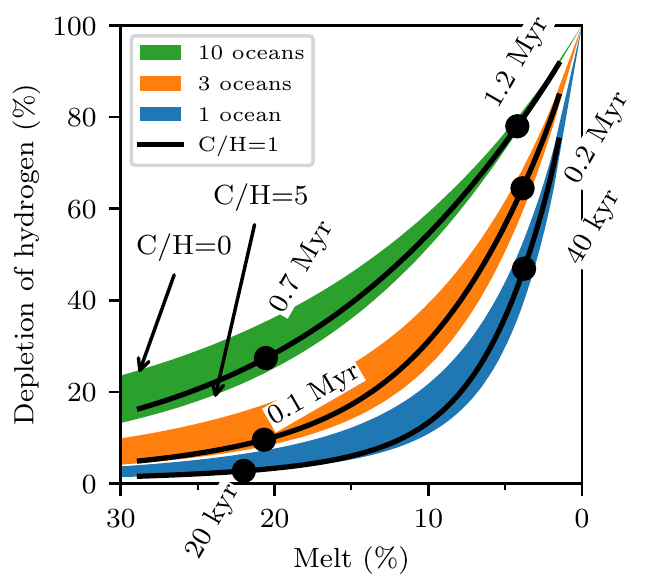}
\caption{Depletion of hydrogen (H) versus melt fraction at $\Delta$IW=0.5 for H budgets of 1, 3, and 10 oceans and different C/H by weight.  For each shaded region, the upper bound is C/H=0, lower bound is C/H=5, and black line is C/H=1.  Symbols indicate times of C/H=1 cases near 20\% melt and 5\% melt.}
\label{fig:ch_ratio_depletion}
\end{figure}

For an inventory of three oceans (middle column, Figure~\ref{fig:ch_ratio}), a larger fraction of the atmosphere consists of H species during the evolution compared to an inventory of one ocean.  This is a direct consequence of the power-law solubility of H$_2$O and the higher amount of H (Appendix~\ref{app:simple}). When no melt remains---as for the end of fractional crystallization---the entire initial inventory of H and C sets the atmospheric composition where equilibrium chemistry continues to dictate the partitioning between reducing and oxidizing species.  Therefore, for fixed C/H at the end of fractional crystallisation, the volume mixing ratios are also set.  If CH$_4$ is insignificant, the final partial pressures ($p_{\rm final}$) scale according to the initial inventory; for example, the final pressures for C/H=0.1 are a factor of three larger for a three-ocean inventory compared to a one-ocean inventory (compare $p_{\rm final}$ in Figure~\ref{fig:ch_ratio}d,e).  For large H inventories, the volume mixing ratio of H species changes less during outgassing because initially more H (compared to C) resides in the atmosphere owing to the power-law solubility of H$_2$O.  The case with a three-ocean budget and C/H=1 (Figure~\ref{fig:ch_ratio}h,~\ref{fig:ch_ratio_partial}e) is similar to the bulk silicate Earth calculated in \cite{SOSSI20}.  At high temperatures, a CO-dominated atmosphere forms with $\sim$200-bar C-bearing species and $\sim$10-bar H-bearing species in which \textit{f}H$_2$ and \textit{f}H$_2$O are subequal, while \textit{f}CO $>$ \textit{f}CO$_2$.

\begin{figure}[tbhp]
\plotone{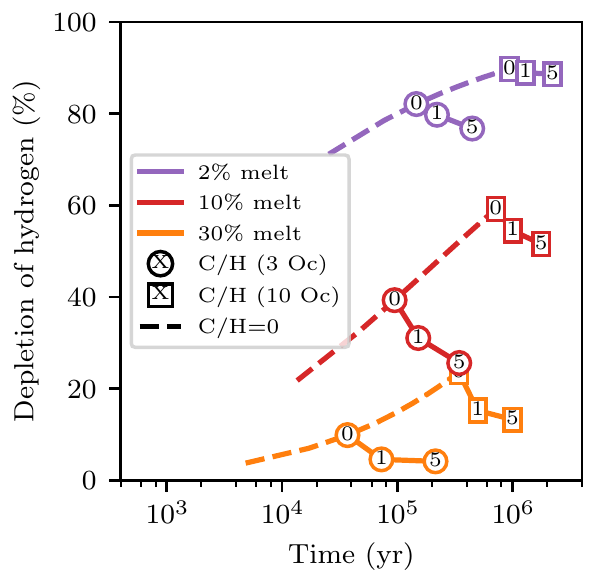}
\caption{Depletion of hydrogen (H) versus time at $\Delta$IW=0.5 for different C/H, where colored lines show constant melt fraction.  Dashed lines show C/H=0 for H budgets from 1 to 10 Earth oceans; these are the solid lines in Figure~\ref{fig:compare_depletion}b.  Departing from the dashed lines, circles show 3 oceans and C/H=X, where X is 0, 1, or 5. Similarly, rectangles show 10 oceans and C/H=X, where X is 0, 1, or 5.}
\label{fig:ch_ratio_depletion_time}
\end{figure}

Figure~\ref{fig:ch_ratio_depletion} summarizes the interior depletion of H for 1--10 oceans and C/H by weight varying from zero to five.  To first order, depletion at a given melt fraction is dictated by the total inventory of H due to the power-law solubility of H$_2$O.  For 10\% melt, this gives rise to around a 40\% increase in depletion as the H inventory increases from 1 to 10 oceans.  A secondary influence on depletion is C/H, where higher C/H suppresses the outgassing and hence depletion of H.  This effect is not connected to the solubility of H$_2$O, but rather is due to the dominant presence of C species in the atmosphere that increases the mean molar mass of the atmosphere relative to H species (Appendix~\ref{app:volmass}).  Figure~\ref{fig:ch_ratio_depletion_time} reveals the influence of C/H on the depletion during magma ocean outgassing and can be compared with Figure~\ref{fig:compare_depletion}.  For a fixed H inventory, increasing C/H prolongs the time taken to reach a given melt fraction since the optical depth of the atmosphere increases. Increasing C/H increases the volume mixing ratio of C species in the atmosphere, and this alone would actually decrease the opacity of the atmosphere since CO has the lowest opacity of all the considered volatiles (Appendix~\ref{sect:opacity}).  However, increasing C/H also increases the total surface pressure substantially because CO$_2$ is relatively insoluble compared to H$_2$O.  Hence, the optical depth increases in response to the larger surface pressure, and this increase predominates over the change in speciation; the net outcome is prolonged cooling for larger C/H.  Figure~\ref{fig:ch_ratio_depletion_time} further demonstrates the trend of Figure~\ref{fig:ch_ratio_depletion}, in which increasing C/H suppresses the outgassing and hence interior depletion of H at a given melt fraction.

\subsection{Oxygen Fugacity}
\begin{figure*}[tbhp]
\centering
\includegraphics[height=0.9\textheight]{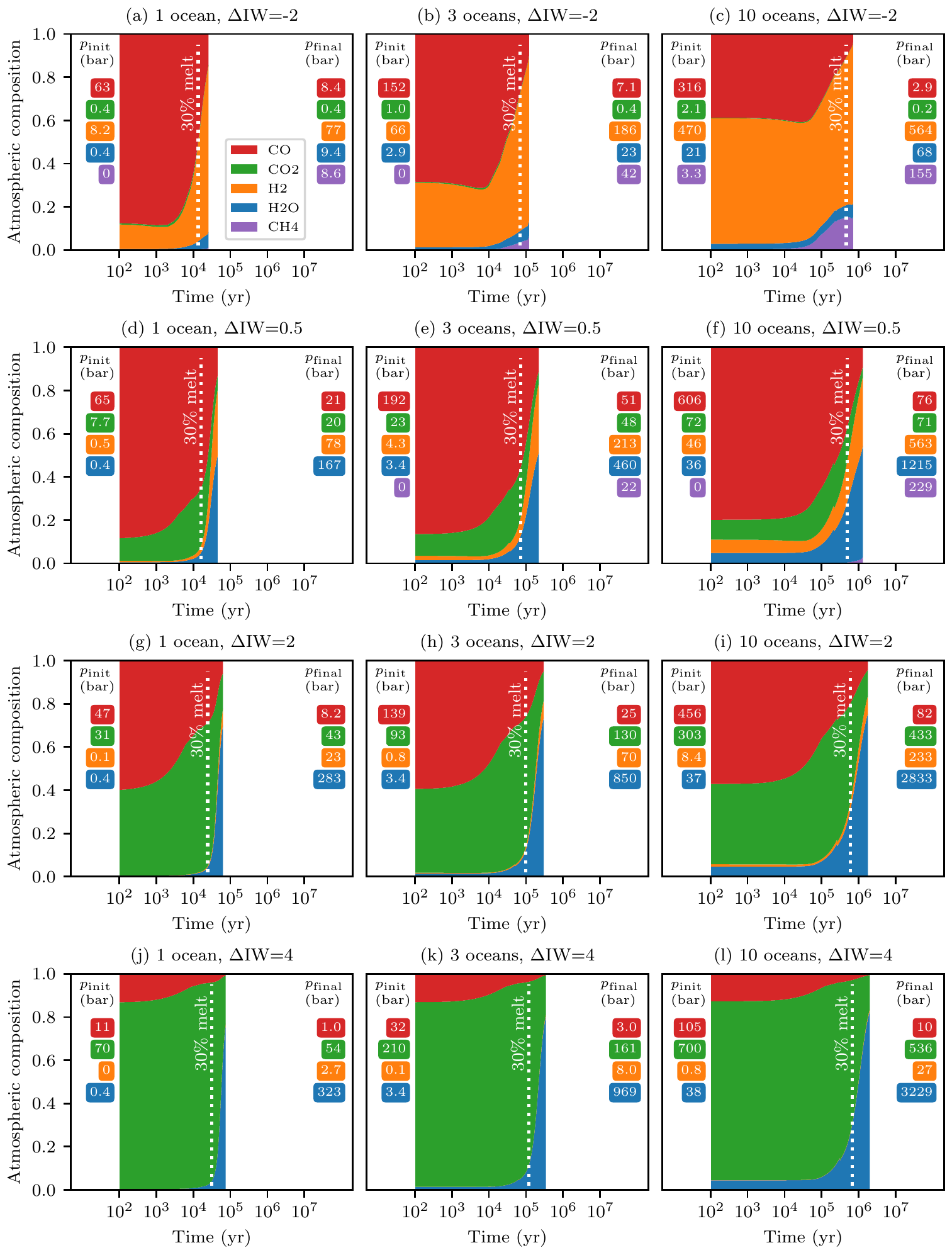}
\caption{Atmospheric composition during magma ocean outgassing for C/H=1 and different $f{\rm O_2}$.  Columns from left to right show hydrogen (H) budgets of 1, 3, and 10 oceans, respectively.  Rows from top to bottom show $f{\rm O_2}$ of $\Delta$IW=$-2$, $\Delta$IW=0.5, $\Delta$IW=2, and $\Delta$IW=4.  Magma ocean lifetime corresponds to the termination of the colored region (at $\approx$ 2\% melt for fractional crystallization, between $10^4$ and $10^7$ yr).  White dotted lines correspond to 30\% melt, which indicates both the duration and atmospheric composition if instead the mantle underwent equilibrium crystallization.  Initial volatile partial pressures ($p_{\rm init}$) are for a completely molten magma ocean and $T_{\rm surf}=2700$ K.  Final partial pressures ($p_{\rm final}$) are for complete solidification and outgassing of all volatiles and $T_{\rm surf}=1400$ K.}
\label{fig:atmosphere_fO2}
\end{figure*}
\begin{figure}[tbhp]
\plotone{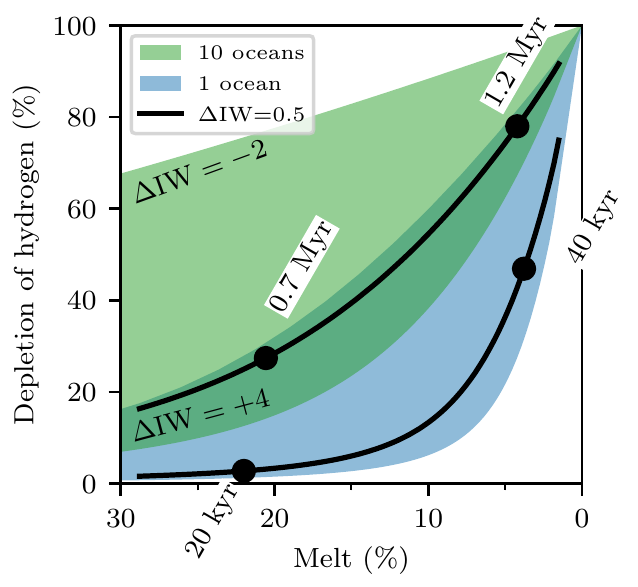}
\caption{Depletion of hydrogen (H) vs. melt fraction for H budgets of 1 and 10 oceans and different $f{\rm O_2}$.  For each shaded region, the upper bound is $\Delta$IW=$-2$, lower bound is $\Delta$IW=4, and black line (same as Figure~\ref{fig:ch_ratio_depletion}) is $\Delta$IW=0.5.  Symbols indicate times of $\Delta$IW=0.5 cases near 20\% and 5\% melt.}
\label{fig:fO2_depletion}
\end{figure}
\begin{figure}[tbhp]
\plotone{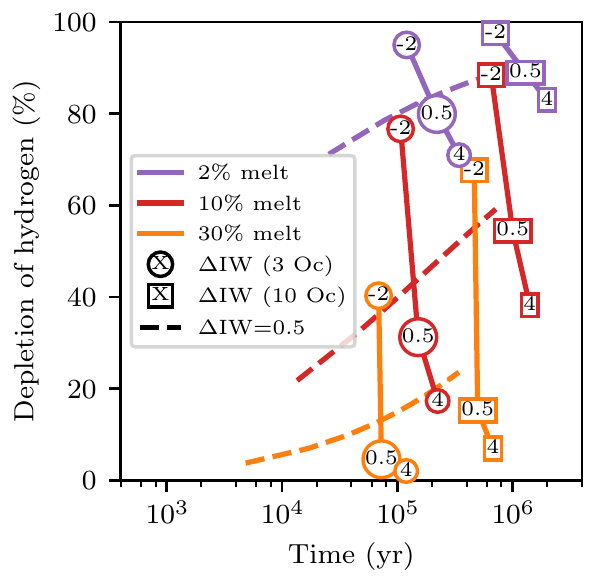}
\caption{Depletion of hydrogen (H) vs. time for different $f{\rm O_2}$, where colored lines show constant melt fraction.  Dashed lines show $\Delta$IW=0.5 and C/H=0 for H budgets from 1 to 10 Earth oceans; these are the solid lines in Figure~\ref{fig:compare_depletion}b.  Departing from the dashed lines, circles show 3 oceans and C/H=1 for $\Delta$IW=X, where X is $-2$, 0.5, or 4. Similarly, rectangles show 10 oceans and C/H=1 for $\Delta$IW=X, where X is $-2$, 0.5, or 4.}
\label{fig:fO2_depletion_time}
\end{figure}

Iron-core formation established the deep Earth's redox state at $\Delta$IW=$-2$ whereas the present-day upper mantle redox is $\Delta$IW=4.  This indicates that mantle redox can vary with both time \citep[e.g.,][]{SG11} and pressure \citep{AFM19}, potentially giving rise to atmospheres that are more reduced or oxidized than our nominal cases at $\Delta$IW=0.5 \citep[e.g.,][]{HMM12}.  Hence, we supplement our calculations at $\Delta$IW=0.5 by additionally considering outgassing scenarios at $\Delta$IW=$-2$, $\Delta$IW=2, and $\Delta$IW=4.  Figure~\ref{fig:atmosphere_fO2} summarizes the results for C/H=1, and detailed figures for all C/H cases at each $f{\rm O_2}$ are presented in Appendix~\ref{app:fO2}.  Increasing the oxygen fugacity from $\Delta$IW=0.5 to $\Delta$IW=4 progressively increases the ratio of oxidized to reduced species in the atmosphere at a given temperature.  Since oxidized species of C and H have a higher molar mass than their reduced counterparts, this also increases the mean molar mass of the atmosphere which influences partial pressures through mass balance (e.g., Equation~\ref{eq:volmass}).  Cooling times are generally extended for oxidized versus reduced atmospheres, due to the higher surface pressure and the intrinsic higher opacity of oxidized species (Appendix~\ref{sect:opacity}).  However, for $\Delta$IW=$-2$ and C/H=5, production of CH$_4$ increases the cooling time to be comparable to an oxidized atmosphere dominated by CO$_2$ and H$_2$O.

Oxidized interiors mitigate the interior depletion of H relative to reducing conditions (e.g., compare $\Delta$IW=$-2$ and $\Delta$IW=4 in Figure~\ref{fig:fO2_depletion} and \ref{fig:fO2_depletion_time}).  At 30\% melt, H depletion is around a factor of 10 larger for $\Delta$IW=$-2$ compared to $\Delta$IW=4.  At 2\% melt, H depletion at $\Delta$IW=$-2$ is approximately 20\% larger than $\Delta$IW=4; we recall that depletion for all cases must reach 100\% when no melt remains.  Depletion of H is largest for $\Delta$IW=$-2$ because it exists mainly as the less soluble H$_2$ and CH$_4$ in the gas, so only a small quantity of H is dissolved in the melt reservoir as H$_2$O.  Whereas increasing C/H mitigates H depletion and extends cooling times, decreasing $f{\rm O_2}$ greatly enhances H depletion (i.e., outgassing) and generally reduces cooling times.  For C/H=1, early atmospheres are almost always dominated by carbon species, as either CO (reduced cases) or CO$_2$ (oxidized cases).  For 10 H oceans and $\Delta$IW=$-2$ the early atmosphere is instead dominated by H$_2$.  Even for the most oxidized scenario when $\Delta$IW=4, H$_2$O is never the dominant species in the atmosphere until the melt fraction drops below about 20\% or even less.  Hence, magma oceans that undergo fractional crystallization are more likely to produce oxidized and H$_2$O-rich atmospheres versus equilibrium crystallization which produces reduced and often CO-rich atmospheres.

\subsection{Atmospheric Escape of H During Outgassing}
\begin{figure*}[tbhp]
\plotone{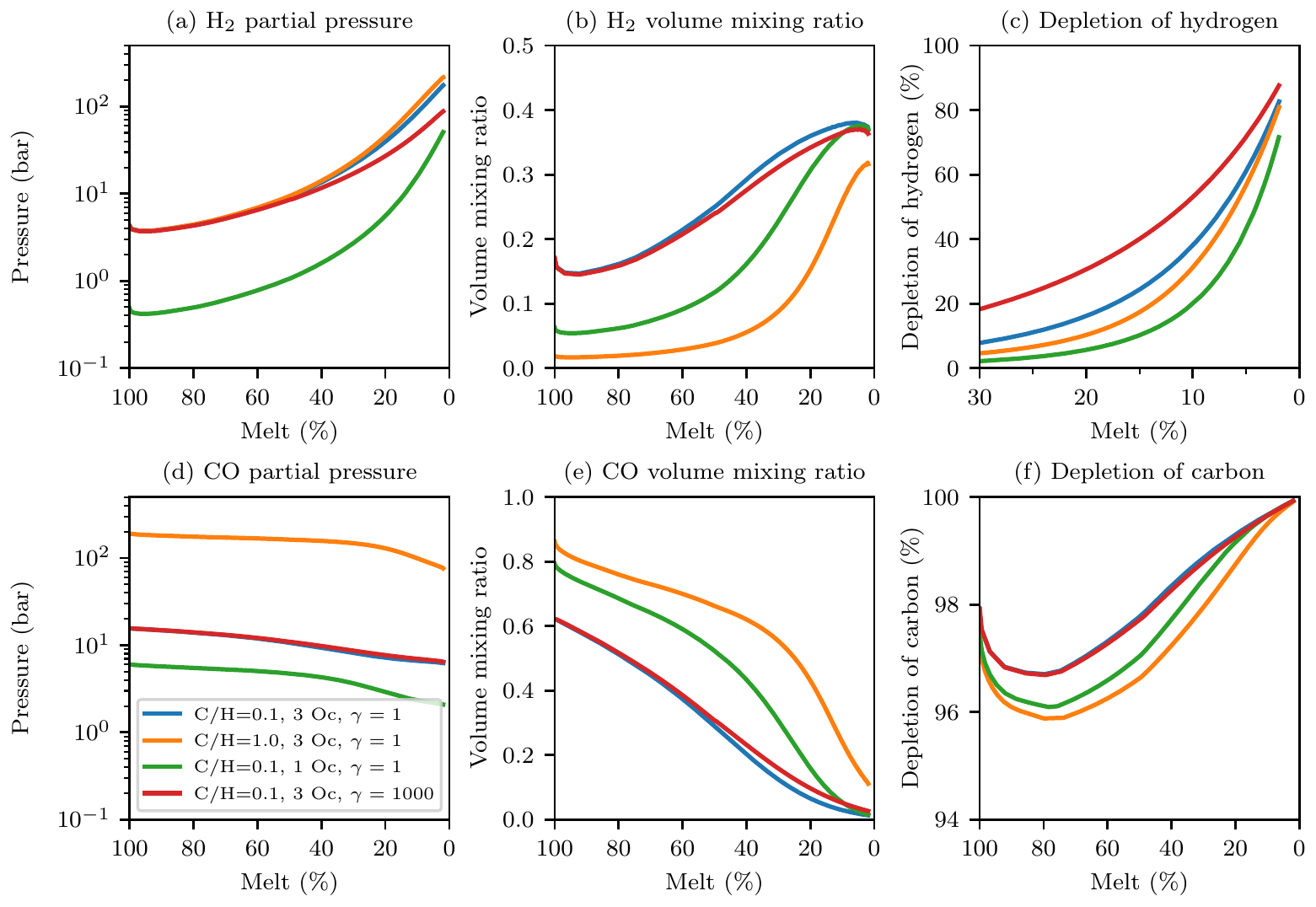}
\caption{Comparison of atmospheric composition and interior depletion at $\Delta$IW=0.5 for (1) C/H=0.1, three H oceans, $\gamma=1$ (2) C/H=1, three H oceans, $\gamma=1$ (3) C/H=0.1, one H ocean, $\gamma=1$ and (4) "Large escape" with C/H=0.1, three H oceans, and an escape prefactor $\gamma=1000$ (Equation~\ref{eq:escape}).  (a, b) H$_2$ partial pressure and volume mixing ratio; (d, e) CO partial pressure and volume mixing ratio; (c, f) interior depletion of H and C, respectively.}
\label{fig:escape}
\end{figure*}
During the magma ocean stage, escape of H$_2$---the lightest atmospheric component---could impact the reservoir evolution of volatiles, so we explore this possibility in Figure~\ref{fig:escape}.  The reference case has C/H=0.1, three H oceans, and unity escape prefactor ($\gamma=1$, Equation~\ref{eq:escape}).  However, an equivalent case with no escape is visually indistinguishable from $\gamma=1$.  This immediately demonstrates that H$_2$ escape due to irradiation of a carbon--hydrogen atmosphere does not appreciably alter the volatile reservoirs during the magma ocean stage for an Earth-sized body at 1 AU, largely because of the magma ocean's short lifetime ($\sim$10$^5$ yr).  Even for a larger H budget of 10 oceans where cooling is relatively prolonged, $\gamma \lesssim100$ has no appreciable impact on the evolution of the volatile reservoirs.  Therefore, we show a case in which escape is increased by setting $\gamma=1000$ ("large escape") to investigate an extreme end-member escape scenario, which may arise, for example, as a result of tidal effects.  For comparison, $\gamma=1000$ gives a mass-loss rate around $2.1\times10^8$ kg s$^{-1}$, which is within the range of rates determined for volatile stripping of the proto-Moon by Earth \citep{CSL21} or for a Mars-sized body near 1 AU undergoing EUV-driven thermal escape \citep{BSL20}.  For large escape the inventory of hydrogen decreases by almost 50\% over the lifetime of the magma ocean of around $10^5$ yr.  By comparison, the magma ocean lifetime for an equivalent case without escape is around $1.5\times10^5$ yr.

Early on (equivalently times at which melt $>$40\%), the H$_2$ partial pressure of all cases with three H oceans follows a comparable trajectory.  At later times (equivalently $<$40\% melt), the cases with three oceans and $\gamma=1$ continue to track a similar trend, although the final H$_2$ partial pressure depends on the initial C/H (e.g., Figure~\ref{fig:ch_ratio}e, h).  For high escape rate, however, the growth of H$_2$ in the atmosphere is mitigated by escape such that the H$_2$ partial pressure is reduced compared to the cases with an unchanging H inventory of three oceans.  Nevertheless, equilibrium chemistry continues to dictate $f{\rm H_2}/f{\rm H_2O}$ so only the magnitudes of the partial pressures of H species are modulated by escape.  The buffering of the atmosphere by the magma ocean presupposes that the loss of H$_2$ is not sufficient to influence the redox state of the magma ocean itself.  In practice, the preferential loss of H$_2$ promotes oxidation of the residual mantle \citep[e.g.,][]{OS19}. The precise amount depends on the abundance of H, but the net effect is to self-arrest loss of H as H$_2$.  This is because as escape proceeds, necessarily more H$_2$ is converted to H$_2$O, given that O-bearing species are heavier and thus escape less readily than H$_2$.  Since for large escape almost 50\% of the initial three-ocean inventory is lost, $f{\rm H_2}$ is steered toward the final pressure (at 0\% melt) of the one-ocean case, which has the same C/H=0.1.  For fixed C/H=0.1, the volume mixing ratio of H$_2$ is the same at 0\% melt regardless of the final H inventory.  However, for large escape, H is lost relative to C, and hence the H$_2$ volume mixing ratio at the end of the magma ocean stage is slightly reduced compared to cases with unchanging C/H=0.1.

Atmospheric escape modulates two controls on outgassing during magma ocean cooling that we have previously investigated: (1) escape decreases the inventory of H and hence modulates the partitioning of H between the interior and atmosphere when melt is present according to the power-law solubility of H$_2$O, and (2) escape increases C/H, resulting in a final atmosphere richer in C compared to H.  For example, the loss of H$_2$ for large escape increases the volume mixing ratio of CO at later time compared to C/H=0.1.  This abides by our previous models that show a larger contribution of carbon species to the atmosphere for a larger C/H.  For large escape, C/H by weight increases from 0.1 initially to 0.2 at the end of the magma ocean stage.  For minimal escape, both decreasing the H inventory and/or increasing C/H suppresses the interior depletion of H at a given melt fraction, relative to the total inventory (Figure~\ref{fig:ch_ratio_depletion}).  However, for large escape, the depletion of H is enhanced relative to other cases (Figure~\ref{fig:escape}).  This demonstrates that the direct loss of H due to escape is a stronger control on its depletion in the magma ocean than the secondary effects on its solubility (through H inventory) and the suppression of H outgassing due to a heavier, carbon-rich atmosphere.  In short, H$_2$ escape driven by irradiation given best estimates for the XUV flux of the young Sun has an insignificant impact on volatile reservoir evolution of an Earth-sized body at 1 AU during the magma ocean stage. However, more extreme escape scenarios can reduce the partial pressure of H species in the atmosphere and drive faster depletion of the interior H reservoir.

\section{Discussion}
\subsection{Atmospheric Composition and Evolution}
\begin{figure*}[tbhp]
\plotone{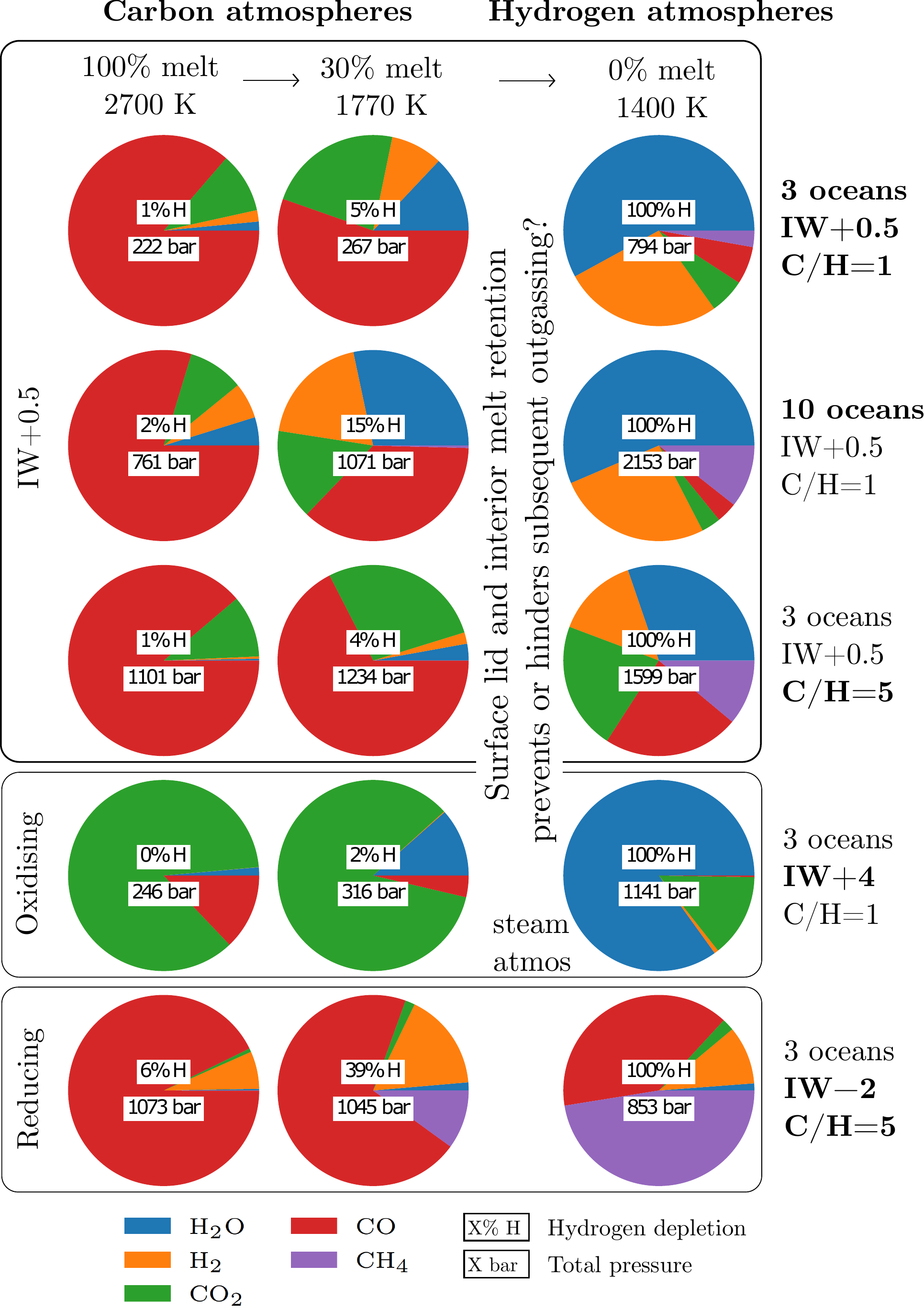}
\caption{Summary of atmospheric composition and hydrogen depletion of the interior during magma ocean cooling at 100\%, 30\%, and 0\% melt, where the surface temperature is given below the melt fraction.  Transition from carbon-dominated to hydrogen-dominated atmospheres depends on surface lid formation and the style of magma ocean crystallization, which may prevent or hinder interior outgassing below 30\% melt fraction.}
\label{fig:summary}
\end{figure*}

For a redox state inferred for the early Earth, the earliest outgassed atmospheres of terrestrial planets are dominantly CO-rich by volume mixing ratio when the mantle is mostly molten (Figure~\ref{fig:summary}).  This is borne out of the equilibrium chemistry of CO--CO$_2$ and the low solubility of C species compared to H species.  With C present, H species can in principle constitute more than 50\% of the atmosphere if the H budget is greater than about 6 Earth oceans owing to the power-law solubility of H$_2$O.  However, this is only possible if C/H by weight is low---around an order of magnitude less than for the bulk silicate Earth.  As C/H increases to unity (Earth-like) and beyond, the atmosphere is CO-rich regardless of the H budget.  Hence, only for C/H$\lesssim0.1$ is an H-rich atmosphere dominant during the early cooling phase, when most of the mantle is molten.  Throughout the magma ocean stage, H$_2$ and H$_2$O have roughly constant mixing ratios owing to the weak temperature dependence of the equilibrium constant (Equation~\ref{eq:h_react}), while $f{\rm H_2}/f{\rm H_2O}$ is approximately unity at $\Delta$IW=0.5 at magmatic temperatures.

For C/H$\gtrsim$1, H species can only begin outgassing in earnest once the melt fraction drops below about 30\%, regardless of the redox state.  However, this is contingent on the surface remaining molten to ensure equilibrium between the interior and atmosphere; this is less likely if the magma ocean undergoes equilibrium crystallization where crystals and melt freeze together and a surface lid forms quickly.  In comparison, fractional crystallization---driven by melt--solid separation---can more likely maintain a molten surface while the deeper mantle crystallizes and exsolves volatiles.  Hence, the transition from a C-rich to H-rich atmosphere depends on the style of crystallization, and crucially the behavior of the mantle as it transitions from mostly molten to mostly solid.  A large steam-dominated atmosphere can only form for a relatively oxidized mantle toward the end of mantle crystallization, although it may persist for a longer duration than the early carbon-rich atmosphere in the absence of an efficient loss mechanism.  An atmosphere that contains initially more H species by volume mixing ratio (C/H$\lesssim0.1$ and particularly for H inventories $>$ 6 oceans) undergoes less modification during outgassing to reach its final volatile mixing ratios when the mantle has fully crystallized.

The chemical boundary layer at the surface of the magma ocean facilitates interior--atmosphere equilibrium, where the equilibrium timescale is rapid relative to a crystallization timescale of around a million years \citep{PSH19}.  For low $f{\rm O_2}$ or small atmospheres, the crystallization timescale decreases by no more than two orders of magnitude ($\sim10^4$ yr) and this decrease would be further mitigated by a greater H$_2$ greenhouse forcing \citep{LBH21}.  Other mechanisms to facilitate equilibrium, such as bubble formation due to volatile supersaturation \citep{ET08}, would also reduce the equilibrium timescale.  Nevertheless, if crystallization proceeds too rapidly compared to the equilibrium timescale, then disequilibrium chemistry may arise between the interior and atmosphere.
\subsection{Hydrogen in the Interior}
For otherwise highly soluble volatiles, such as H$_2$O, to exist in the atmosphere, the magma ocean must have crystallized below 30\% melt fraction before a surface lid could form and persist at the surface.  Otherwise, thermodynamic communication between the interior and atmosphere is broken, and dissolved volatiles remain trapped in the mantle.  Below 30\% melt fraction, continued outgassing of dissolved volatiles requires that significant solidification occurs through melt percolation and solid compaction before the rheological front reaches the surface or a quench crust forms (fractional crystallization).  In an end-member scenario, this can enable near-complete outgassing of all volatiles, which has received the most attention in the literature.  At a given melt fraction, we find that a larger fraction of H can be retained in the interior for smaller total inventories of H, larger C/H by weight, and more oxidized interiors.  Although C does not influence H solubility directly, it impacts H retention since it suppresses H outgassing through its influence on the mean molar mass of the atmosphere \citep{BKW19}.

Equilibrium crystallization results in a substantial reservoir of melt and hence volatiles trapped in the interior when the surface forms a lid.  This later possibility has received less attention in the context of volatile evolution during the magma ocean stage, but it provides a mechanism to safely harbor a significant quantity of volatiles in the mantle to protect against loss from atmospheric escape and impacts.  Hence, the high solubility of H$_2$O in silicate melt may be a crucial property of this life-supporting molecule that enables it to survive during the violent early years of a terrestrial planet's life.

Volatile retention would be further enhanced if a melt--crystal density crossover enabled the formation of a voluminous basal magma ocean \citep{LHC07,CHN19} or if melt is captured as the rheological front advances through the mantle \citep{HH17}.  Trapped melt, and hence soluble volatiles, could then be sequestered in the deep mantle by Rayleigh-Taylor instability \citep{MTS17,MK21}.  Moreover, we assume no solubility of H in crystallizing minerals, whereas experimental data indicate significant quantities of water may be stored in ringwoodite \citep{FK20}.  In short, there are additional processes not included in our models that further conspire against the complete outgassing of soluble volatiles, namely, water, during the magma ocean stage.  Furthermore, estimates of the volatile concentration in the bulk silicate Earth are incompatible with complete outgassing \citep[e.g.,][]{HH17}.

Therefore, complete outgassing of soluble volatiles---frequently an outcome of magma ocean models---only occurs with fractional crystallization if the aforementioned processes are absent or inefficient.  Furthermore, it requires that a surface lid or quench crust, if present, does not hinder volatile outgassing as fractional crystallization proceeds.  Hence, at the end of the magma ocean stage it is reasonable to expect that a trapped reservoir of H$_2$O in the interior will interact with the surface environment and atmosphere.  This could occur owing to post-magma ocean cumulate overturn or owing to processes operating over geological timescales (millions to billions of years).  The detailed chemical and physical processes governing the formation and sustenance of cumulate layers and a surface lid in a dynamic magma ocean require further investigation; external influences such as projectile bombardment could also stifle lid formation \citep[e.g.,][]{PJE18}.

For an Earth-like planet orbiting a Sun-like star at 1 AU during its magma ocean stage, we find that atmospheric escape of H$_2$ due to irradiation (energy and diffusion limited) does not significantly impact the evolution of volatile reservoirs, in agreement with \cite{HAG13} and \cite{KOG20}.  This is because escape rates are sufficiently low and the magma ocean duration is sufficiently short (at most a few million years).  In future work, a feedback to probe is that loss of H$_2$ increases $f{\rm O_2}$ slightly, causing more H$_2$ to convert to H$_2$O and thereby partly self-arresting the loss process.  Furthermore, we have not considered photodissociation of H$_2$O, but this could interplay with the included geochemical reactions that depend on mantle redox.

\subsection{Crystallisation Style}
Based on energetic considerations of convection versus gravitational settling, \cite{SS93} propose a critical crystal size of 1 cm above which fractional crystallization is inevitable.  Furthermore, they show that the critical crystal size is weakly dependent on crystal fraction and therefore depth.  Hence, their results are compatible with our fractional crystallization models with a constant crystal size of 5 cm.  In detail, gravitational settling in our models becomes dominant at the rheological transition owing to the substantial decrease of turbulence that enabled crystals to remain mostly in suspension.  It is reasonable to assume that crystals are suspended at high melt fraction \citep{TM90}, although they could begin to settle and thereby initiate fractional crystallization before the rheological transition is obtained \citep{PCT20}, assuming inefficient re-entrainment \citep{SOS93} or a high planetary rotation rate \citep{MH19}.  Settling also depends on mineral buoyancy, and if the minerals float rather than sink, this could hinder interior--atmosphere communication.  For example, preferential partitioning of Fe into melt during crystallization continuously changes the level of neutral buoyancy between crystals and melt \citep{CHN19}.  In contrast to fractional crystallization, equilibrium crystallization prevents differentiation of the mantle and occurs for a smaller crystal size of 1 mm \citep{SS293}.

Most previous dynamic models of magma ocean cooling do not consider melt--solid separation \citep{LMC13,HAG13,SWB16,NKT19}; rather, the interior is assumed to always be adiabatic.  Yet some of these previous models report agreement with the solidification timescale derived from geochemical models that explicitly consider fractional crystallization.   The previous models implicitly assume that although the deep mantle reaches the rheological transition first, it continues cooling as efficiently by convection as the uppermost mantle; this allows the melt fraction to continue decreasing at a similar rate.  However, this is unlikely given the rheological behavior of a melt--solid aggregate \citep[e.g.,][]{CCB09}, which at the rheological transition predicts a large reduction in the convective velocity and rapid melt--solid separation \citep{ABE93}.  Hence, somewhat coincidentally, previous models decrease the melt reservoir and outgas volatiles at a similar rate to geochemical fractional crystallization models, even though the key ingredient to justify fractional crystallisation (i.e., melt--solid separation) is not included.
\subsection{Graphite, Diamond, and Water Precipitation}
\label{sect:precipitation}
Our model assumes that all volatile species dissolved at the surface of the magma ocean continue to remain so, irrespective of pressure (depth of the magma ocean) and temperature (cooling of the magma ocean).  Phase transformations involving these volatile elements are likely to occur as pressure and temperature change \citep[e.g.,][]{HMM12}. However, experimental data for the speciation and partitioning of H-, C-, and O-bearing volatile species rarely exceed 7 GPa (Section 2.2), which precludes a holistic, bottom-up model of magma ocean crystallization.  Nevertheless, we now systematically investigate possible phase transformations of H and C to identify scenarios under which the pressures of outgassed atmospheres could diverge from those calculated by our model.

\begin{figure}[tbhp]
\plotone{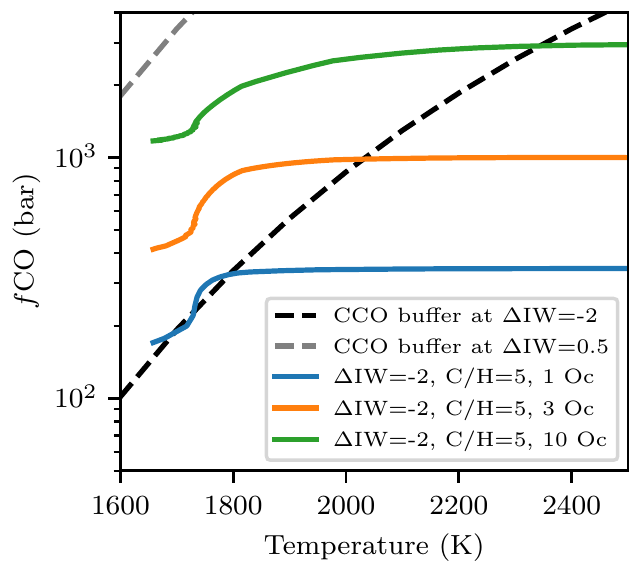}
\caption{The fugacity of CO in equilibrium with graphite (Equation~\ref{eq:cco}) assuming ideal gas behavior as a function of temperature at $\Delta$IW=$-2$ (dashed black line) and $\Delta$IW=0.5 (dashed gray line).  All cases at $\Delta$IW=$-2$ and C/H=5 (colored lines) intercept the CCO buffer at $\Delta$IW=$-2$, indicating graphite precipitation.  Only cases that intercept the CCO buffer are plotted.}
\label{fig:cco}
\end{figure}

Reducing conditions may induce graphite or diamond precipitation \citep{HMM12,TOT13,KG19}. The precipitation of graphite occurs if the CO fugacity of the atmosphere exceeds that defined by the graphite--carbon monoxide (CCO) buffer:
\begin{equation}
\begin{aligned}
\rm{C(s)} + 0.5 \rm{O_2 (g)} &= \rm{CO (g)}, \\
\log_{10}{\rm K_{eq}} &= \frac{6254}{T} + 4.334,
\end{aligned}
\label{eq:cco}
\end{equation}
where $\mathrm{K_{eq}}$ as a function of temperature is provided by the JANAF database \citep{JANAF}.  Equation~\ref{eq:cco} indicates that $\log_{10}f$CO in equilibrium with graphite decreases proportional to $0.5\log_{10}f$O$_2$ and increases with temperature. Thus, graphite precipitation is favored under more reducing conditions and lower temperatures (Figure~\ref{fig:cco}).  For our cases at $f\rm{O_2}$s of $\Delta$IW=0.5 or more oxidized, graphite precipitation would not occur because $f$CO never exceeds that defined by the CCO buffer. This suggests that it is unlikely that a terrestrial magma ocean was saturated in graphite at high melt fractions ($>30$\%).  However, it may occur for most atmospheres as the surface cools below the temperature at which a surface lid forms around 1650~K \citep{SOSSI20}.  At this stage, however, the atmosphere and mantle are not necessarily in equilibrium and the atmosphere can evolve as a near-closed system.

For a magma ocean more reduced than IW, the modeled cases at $f\rm{O_2}$ of $\Delta$IW=$-2$ and C/H=5 show that graphite precipitation would occur in a cooling atmosphere, even accounting for the production of CH$_4$ that buffers $f$CO (Figure~\ref{fig:cco}). Specifically, higher initial C budgets result in higher $f$CO, where, for C/H=5, the scenario with 10 oceans crosses the CCO buffer at $\sim$2300~K, while it does so at $\sim$2000 K for five oceans. As such, the quoted fugacities of carbon-bearing species for these two scenarios are upper limits. Cases at C/H=1 and $\Delta$IW=$-2$ do not result in graphite saturation at any temperature, irrespective of the number of oceans, owing to the lower $f$CO.
We presume that the magma ocean is well mixed, such that it remains isochemical throughout its depth. However, the saturation of graphite in equilibrium with silicate melt depends on pressure and temperature \citep[][]{DCS13,SHW14,CDD14}.  As such, chemical gradients could arise from its precipitation during magma ocean cooling and/or at depth followed by its isolation from the melt \citep{HMM12}.  To assess the potential effect of graphite/diamond precipitation on the calculated fugacities of carbon-bearing species, we compare the C contents dissolved in the magma ocean predicted by our models with the graphite/diamond precipitation curve. Above the IW buffer, C dissolves predominantly as CO$_3^{2-}$ in mafic and ultramafic melts \citep{AHS15,DDT17}.  \cite{HPG92} devised an expression to determine CO$_3^{2-}$ solubility at graphite saturation.  To calculate whether graphite is expected to precipitate from the silicate melt in our simulations, we adopt the calibration of \cite{DDT17} extrapolated to peridotite compositions. At constant pressure (1 GPa) and for the temperature range of cooling considered in our models (1650 to 2500 K), little or no C precipitation occurs in the magma ocean under any of the scenarios modeled at IW+0.5 and above.  At 1650 K and 1 GPa, the model of \cite{DDT17} predicts that $\sim$350 ppmw CO$_2$ is dissolved in a peridotite melt or 100 ppmw for a komatiitic melt (expected for an evolved magma ocean) at graphite saturation, compared to $\sim$300 ppmw in the melt in the most extreme, C/H=5, 10-ocean case.

Below the IW buffer, CO$_3^{2-}$ is no longer the prevailing melt species of dissolved carbon  \citep[instead, carbon likely occurs as some CO-bearing molecule,][]{WRJ13,DHJ19}; therefore other expressions for the prediction of C solubility at graphite saturation need to be considered.  Here we adopt the calibration of \cite{YNN19} that links CO dissolved in the melt to its fugacity, $f\rm{CO}$. We fix $f\rm{CO}$ as a function of pressure in equilibrium with graphite at IW$-2$ using thermodynamic data and the equation of state of \cite{JO94}. At 2 GPa, this yields an $f\rm{CO}$ of $\sim$8.3 GPa, which, using the expression of \cite{YNN19}, leads to 500 ppmw of dissolved CO. These estimates far exceed those derived by solely considering CO$_3^{2-}$ as the dissolved species \citep[$\sim$2 ppmw,][]{DDT17}. Given that dissolved C contents reach 20 ppmw at most in our simulations at IW$-2$, this implies that graphite and diamond should not precipitate at any stage in the magma ocean, even under reducing conditions.

Another important phase change can occur in cooling atmospheres, namely, that between liquid water and steam:
\begin{equation}
{\rm H_2O(l)} = {\rm H_2O(g)}.
\label{eq:h2o-cond}
\end{equation}
Its equilibrium constant as a function of temperature indicates that, at 1400 K, H$_2$O condenses at about 600-bar $f$H$_2$O. This phase transformation is neglected in our simulations primarily because these pressures and temperatures exceed the critical point of water, which is 650 K and 220 bars.  Hence at 600-bar $f$H$_2$O the atmosphere is a supercritical fluid rather than an ideal gas as modeled herein.  Moreover, the mixing of CO$_2$ and H$_2$O is nonideal, such that departures from ideal gas behavior are expected in these solutions \citep{DMW92,FW97}.

\subsection{Meteoritic Degassing}
Theoretical calculations \citep{SF10} and outgassing experiments \citep{TTS21} predict the composition of gases evolved from the degassing of meteoritic materials, some of which may be representative of planetary building blocks.  By examining equilibrium between gas and condensed phases, thermodynamic models find that H$_2$ and CO gases form from reduced chondrites (such as EH) at high temperature \citep{SF10}.  Only for the most oxidizing materials (CI chondritic composition) is H$_2$O predicted to be a major constituent of the gas mixture.  Since water is only an important gaseous species during degassing of one particular class of chondrite, this calls into question the applicability of steam atmospheres to the evolution of terrestrial planets at large.

Although instructive, directly correlating gases produced from meteorites to the composition of planetary atmospheres is complicated for several reasons.  Meteorites are imperfect analogs of the material that went on to form the terrestrial planets \citep{SOSSI21}, particularly as regards volatile elements due to the ubiquitous alteration processes occurring on their parent body \citep{BRE06}, as well as thermal metamorphism and the likelihood that planetesimals are already differentiated when they accrete.  More importantly, the likely presence of a deep magma ocean during and immediately after accretion would have modified the budgets of elements available to degas at the planetary surface \citep[e.g.,][]{GDF20,DHJ19}.  As shown in our work, equilibrium reactions between silicate melt and volatiles in the gas phase influence the partitioning of gas species between the atmosphere and magma ocean according to their solubility.  Furthermore, some volatiles (including H and C) may be permanently sequestered into an iron core owing to their siderophile nature \citep[e.g.,][]{TSH21}.

Therefore, connecting gas mixtures to atmospheric composition necessitates an assumption that the starting bulk composition---usually derived from chondrites--adequately captures elemental abundances in the atmosphere during and after the magma ocean phase.  This assumes that volatiles did not experience significant partitioning due to magma ocean and core formation processes and that atmosphere formation proceeds by the breakdown of solid mineral assemblages (i.e., degassing occurs subsolidus).  Solid bodies cannot efficiently replenish surface material from deeper regions, whereas for molten bodies the entire melt mass can potentially communicate with the atmosphere, subject to the interior--atmosphere equilibration timescale \citep[e.g.,][]{PSH19}.  Alternatively, partitioning can be permitted if ongoing impact-induced outgassing of chondritic materials after a surface lid or quench crust forms (e.g. in a late-veneer scenario or during the late heavy bombardment) is sufficient to displace or dilute any previously equilibrated atmosphere \citep{ZLC20}.

Extraction of volatiles to solids in the core or mantle locks them away from outgassing to the atmosphere over a short timescale, and perhaps indefinitely.  By contrast, volatiles partitioned in melt are readily available to outgas during mantle crystallization as long as the interior and atmosphere remain in thermodynamic communication (e.g., for fractional crystallization).  Accounting for thermodynamic equilibrium with silicate melt, we demonstrate that the earliest atmosphere evolves in both mass and composition during magma ocean cooling.  An atmosphere can transition from dominantly CO-rich ("reducing") to H$_2$O-rich ("oxidizing") owing to solubility and redox reactions as the magma ocean cools (Figure~\ref{fig:summary}).  Therefore, the solubility of volatiles in silicate melt prevents drawing a simple connection between the composition of planetary building blocks and the earliest atmospheres of terrestrial planets.

\section{Summary and Conclusions}
The investigation of catastrophic outgassing of CO$_2$ and H$_2$O from terrestrial magma oceans---as well as their influence on the nature of early atmospheres around rocky planets---was originally motivated by the degassing of hydrated minerals and oxidized chondritic materials.  However, for redox conditions appropriate for the early Earth, we find that CO is usually the dominant atmospheric species during the early molten stage, where the melt fraction is greater than 30\% (Figure.~\ref{fig:summary}).  This is due to the relatively high solubility of hydrogen species (H$_2$O) compared to carbon species (CO$_2$), as well as redox reactions that govern CO--CO$_2$, H$_2$--H$_2$O, and CO$_2$--H$_2$--CH$_4$ equilibria.  Only when C/H by weight is small (C/H$<$0.1) and the H budget large ($>$6 oceans) can H species constitute more than 50\% of the atmosphere, with H$_2$ and H$_2$O in approximately equal abundance (Figure~\ref{fig:ch_ratio}).  For more oxidizing conditions, early atmospheres are dominated by CO$_2$ with H$_2$O a minor species (Figure~\ref{fig:atmosphere_fO2}).

For the late molten stage ($<30\%$ melt), more H is retained (relative to the total budget) for smaller H budgets, larger C/H, and more oxidizing conditions (Figure~\ref{fig:summary}).  At this stage, continued outgassing driven by dissolution equilibrium requires that the surface remains molten and a persistent lid does not form; this requirement can be satisfied if a magma ocean undergoes fractional crystallization.  However, for magma oceans that freeze more quickly by equilibrium crystallization, formation of a surface lid around 30\% melt can prevent subsequent outgassing of highly soluble volatiles such as H$_2$O.  This would leave the atmosphere as carbon/CO dominated by stifling the formation of a steam atmosphere.  Methane only forms for reduced conditions ($\Delta$IW$\lesssim$0.5) when pressures are sufficiently high or temperatures sufficiently low; otherwise, CH$_4$ is absent.  Graphite precipitation from the atmosphere during magma ocean cooling is expected only for very high C budgets (higher than those anticipated for Earth) and for $\Delta$IW$\lesssim$0.5.  Nevertheless, for all cases, graphite in the atmosphere may precipitate with continued cooling.

Complete outgassing of volatiles during magma ocean solidification can arise as a result of fractional crystallisation.  However, by additionally considering equilibrium crystallization and other processes, we expect that a substantial reservoir of H could remain in planetary mantles and outgas over geological timescales, potentially impacting the depth of surface oceans and the ability to desiccate planets.  In this case, hydrogen species would only play a minor role in determining atmospheric opacity or modulating atmospheric escape during the early magma ocean; rather, the behavior of carbon species is dominant.  The style of magma ocean crystallization---fractional or equilibrium---therefore controls the composition and mass of early atmospheres and the efficiency of volatile delivery to the planetary atmosphere and surface.  Ultimately, the high solubility of H$_2$O in magma oceans may enable its safe storage during the tumultuous phase of planet formation.

\section{Data Availability}
All of the data generated as part of this study can be obtained by contacting Dan J. Bower.

\begin{acknowledgements}
DJB acknowledges Swiss National Science Foundation (SNSF) Ambizione Grant 173992. KH is supported by the European Research Council via Consolidator Grant ERC-2017-CoG-771620-EXOKLEIN.  PAS acknowledges Swiss National Science Foundation (SNSF) Ambizione Grant 180025. PS acknowledges financial support from the Swiss University Conference and the Swiss Council of Federal Institutes of Technology through the Platform for Advanced Scientific Computing (PASC) program.   This research was partly inspired by discussions and interactions within the framework of the National Center for Competence in Research (NCCR) PlanetS supported by the SNSF.  The calculations were performed on UBELIX (\url{http://www.id.unibe.ch/hpc}), the HPC cluster at the University of Bern.  We thank S. Grimm and D. Kitzmann for computing Rosseland mean opacities, and A. Wolf, M. Tian, and T. Lichtenberg for discussions and feedback on the manuscript.  Three anonymous referees provided comments that further enhanced the manuscript.
\end{acknowledgements}

%




\software{Simulating Planetary Interior Dynamics with Extreme Rheology (SPIDER) \citep{BSW18,BKW19} is open source and hosted at \url{https://github.com/djbower/spider}.  SPIDER Version 0.2.1 was used in this study and is also available on Zenodo \citep{SPIDERzenodo}.}



\clearpage

\appendix

\renewcommand{\thetable}{\Alph{section}} 

\twocolumngrid

\section{Opacities}
\label{sect:opacity}
We computed Rosseland mean opacities at 1 bar between 1700 and 2700 K for H$_2$O, H$_2$, CO$_2$, CO, and CH$_4$ using \texttt{Helios-k} \citep{GMK21,GH15}.  The opacity (Rosseland mean mass absorption coefficient) for H$_2$O using the full BT2 list \citep{BTH06} is $1.5 \times 10^{-2}$ m$^2$ kg$^{-1}$ at 1700 K and $5.0 \times 10^{-4}$ m$^2$ kg$^{-1}$ at 2700 K; the opacity at 1800 K compares favorably with $10^{-2}$ m$^2$ kg$^{-1}$ which has been used extensively in previous magma ocean studies.  For simplicity we also adopt a constant value (i.e., independent of temperature) of $10^{-2}$ m$^2$ kg$^{-1}$.  The opacity of H$_2$ is determined by collision-induced absorption, and using HITRAN \citep{KGA19,AFL11} it is essentially constant ($5 \times 10^{-5}$ m$^2$ kg$^{-1}$) from 2700 to 1700 K.

For CO$_2$, HITEMP \citep{RGB10} provided the line list, and we determined an opacity of $1.5 \times 10^{-4}$ m$^2$ kg$^{-1}$ at 1700 K and $3 \times 10^{-5}$ m$^2$ kg$^{-1}$ at 2700 K.  Hence the CO$_2$ opacity at 1700 K is compatible with $10^{-4}$ m$^2$ kg$^{-1}$, which is the lowest opacity considered for CO$_2$ in previous magma ocean models.  To be consistent with our selection of the H$_2$O opacity around 1800 K, we similarly adopt a constant CO$_2$ opacity of $10^{-4}$ m$^2$ kg$^{-1}$.  For CO we determined an opacity of $1.2 \times 10^{-5}$ m$^2$ kg$^{-1}$  at 1700 K and $3 \times 10^{-6}$ m$^2$ kg$^{-1}$ at 2700 K using HITEMP \citep{LGR15}.  Again, we selected an opacity of $10^{-5}$ m$^2$ kg$^{-1}$ based on the value at 1800 K.  We computed a CH$_4$ opacity of $10^{-2}$ m$^2$ kg$^{-1}$ at 1800 K based on the line lists of \cite{YTB13,YT14}.  The opacities of species at 1 bar are reference opacities that are used to compute a pressure-dependent opacity \citep[Equation~A22,][]{AM85}.

\section{Melt--Solid Separation}
\label{sect:sep}
The fluxes to describe energy transport in a planetary mantle (which is molten, partially molten, or solid) are described in detail in \cite{BSW18,ABE95}.  However, we recap the physics underpinning melt--solid separation owing to its importance for determining whether a mantle undergoes fractional or equilibrium crystallization.  At high melt fraction, the settling or flotation of crystals within a magma ocean is determined by Stokes's law:

\begin{equation}
    u_m-u_s = \frac{2 a^2 g (\rho_m-\rho_s)}{9 \eta_m},
\end{equation}
where $u$ is velocity (positive radially outward), $a$ crystal size, $g$ gravity (negative), $\rho$ density, $\eta$ viscosity, and subscripts $m$ and $s$ denote melt and solid, respectively.  Interaction amongst crystals is not considered, and therefore free settling or flotation of crystals is an upper limit on the efficiency of melt--solid separation at high melt fraction.  At low melt fraction, separation occurs via Darcy flow:

\begin{equation}
    u_m-u_s = \frac{k_p g(\rho_m-\rho_s)}{\phi_p \eta_m},
\end{equation}
where $k_p$ is permeability and $\phi_p$ porosity (synonymous with the volume fraction of melt).  The Rumpf--Gupte permeability law is appropriate at intermediate porosity \citep{RG71},

\begin{equation}
    k_{rg} = \frac{a^2 \phi_p^{5.5}}{1.4},
\end{equation}
and the Blake--Kozeny--Carman permeability law at low porosity,

\begin{equation}
    k_{bkc} = \frac{a^2 \phi_p^3}{1000(1-\phi_p)^2}.
\end{equation}
The constants in these permeability laws are constrained by experiments \citep[see][for discussion]{M84,ABE95}.  The flow laws can be represented by a single description \citep[][]{ABE95}:

\begin{equation}
u_m-u_s = \frac{a^2 g (\rho_m-\rho_s)}{\eta_m} f,
\label{eq:grav_du}
\end{equation}
\begin{subnumcases}{f = \label{eq:grav_f}}
    \dfrac{2}{9} & $\text{for } 0.77 \leq \phi_p$, \label{eq:grav_stokes}\\ 
    \dfrac{k_{rg}}{a^2 \phi_p} & $\text{for } 0.08 < \phi_p < 0.77$, \label{eq:grav_rg}\\
    \dfrac{k_{bkc}}{a^2 \phi_p} & $\text{for } \phi_p \leq 0.08$, \label{eq:grav_bkc} 
\end{subnumcases}
where the range of applicability of each flow law ensures that $f$ is a continuous function of porosity.  Through consideration of the local barycentric velocity, the mass flux of melt is

\begin{equation}
    J_m = \rho \phi (1-\phi) (u_m-u_s),
    \label{eq:grav_Jm}
\end{equation}
where $\phi$ is melt fraction (the mass fraction of melt).  Equation~\ref{eq:grav_Jm} naturally satisfies the requirement that $J_m=0$ for $\phi=0$ and $\phi=1$.  The energy flux of melt--solid separation is therefore

\begin{equation}
    F_{\rm grav} = J_m T_{\rm fus} \Delta S_{\rm fus},
\end{equation}
where $T_{\rm fus}$ is the temperature at 50\% melt fraction and $\Delta S_{\rm fus}$ is the entropy of fusion, both of which are functions of pressure according to the melting curves.  Equation~\ref{eq:grav_du} reveals the common factors appearing in all of the flow laws and emphasizes the importance of crystal size owing to its squared dependence: larger crystals enhance melt--solid separation.  Since we consider a single-component mantle (MgSiO$_3$), everywhere $\rho_m-\rho_s<0$ and hence crystals always sink and melt always drains upward to the surface.
\section{Magma ocean and atmosphere equilibrium}
\label{app:volatiles}
\subsection{Volatile mass balance and evolution}
\label{app:volmass}
We extended \cite{BKW19} to express the mass balance for a volatile $v$ in a chemically reactive environment with atmospheric escape:

\begin{align}
X_v k_v M^s + X_v M^l + 4 \pi R_p^2 \left( \frac{\mu_v}{\bar{\mu}} \right) \frac{p_v}{|g|} + m_v^e + \sum_{w=1}^r m_v^w \nonumber \\
= X_v^0 M^m,
\label{eq:volmass}
\end{align}
where:
\begin{itemize}
\item $X_v$ is the volatile abundance in the pure melt;
\item $k_v$ is the distribution coefficient between solid and melt;
\item $M^s$ is the mantle mass of solid;
\item $M^l$ is the mantle mass of melt;
\item $M^m = M^s + M^l$ is the total mantle mass, which is constant;
\item $R_p$ is the planetary surface radius;
\item $g$ is the surface gravity;
\item $\mu_v$ is the molar mass of the volatile;
\item $\bar{\mu}$ is the mean molar mass of the atmosphere;
\item $p_v$ is the surface partial pressure of the volatile; and
\item $X_v^0$ is the initial total volatile abundance relative to the total mantle mass.
\end{itemize}
A solubility law relates the volatile concentration in the melt ($X_v$) to the partial pressure of the volatile in the atmosphere ($p_v$).  The first three terms on the left-hand side of Equation~\ref{eq:volmass} represent the mass of the volatile stored in the solid mantle, the molten mantle, and the atmosphere.  The final two terms on the left-hand side are fictitious reservoirs used to account for mass loss due to escape ($m_v^e$; Appendix~\ref{app:escape}), and mass transfer due to participation of this volatile in reaction $w$ ($m_v^w$, Appendix~\ref{app:chemreact}).  The time derivative of Equation~\ref{eq:volmass} derives the evolution equation \citep[cf., Equation~A4,][]{BKW19}:
 
\begin{align}
&\frac{d X_v}{d t} \left(k_v M^m + (1-k_v) M^l \right)
+ X_v (1-k_v) \frac{d M^l}{d t} \nonumber \\
& + \frac{4 \pi R_p^2 \mu_v}{|g|}
\left[\frac{1}{\bar{\mu}} \frac{d p_v}{d t}
-\frac{p_v}{\bar{\mu}^2} \sum_{w=1}^n \frac{\mu_w}{P_s} \left( \frac{d p_w}{d t} - \frac{p_w}{P_s} \frac{d P_s}{d t} \right)
\right] \nonumber \\
& + \frac{dm_v^e}{dt} + \sum_{w=1}^r \frac{dm_v^w}{dt} = 0,
\label{eq:volmassdt}
\end{align}
where $P_s$ is the total surface pressure, which is equal to the summation of the partial pressures of all volatiles according to Dalton's law.  The derivative of $X_v$ with respect to $t$ can be determined from the solubility law relevant for volatile $v$, which for power-law solubility (Equation~\ref{eq:sol}) is
\begin{equation}
\frac{dX_v}{dt} = \frac{dX_v}{dp_v} \frac{dp_v}{dt} = \left( \frac{\alpha_v {p_v} ^ {1/\beta_v-1}}{\beta_v} \right)  \frac{dp_v}{dt}.
\end{equation}
Similarly, $X_v$ can be eliminated from Equation~\ref{eq:volmassdt} using the solubility law.  This results in a system of nonlinear equations that can be solved simultaneously to determine $dp_v/dt$ for each volatile $v$.  Solving for $dp_v/dt$ rather than $dX_v/dt$ enables us to treat the case of zero solubility.  Oxygen fugacity $f{\rm O_2}$ (Equation~\ref{eq:fO2}) depends on surface temperature, so oxygen is not explicitly tracked by mass balance as with the other volatiles.
\subsection{Mass loss due to atmospheric escape}
\label{app:escape}
The escape flux of H$_2$ (Equation~\ref{eq:escape}) is used to determine a mass-loss rate of H$_2$ \citep[e.g.,][]{KOG20}: 

\begin{equation}
\frac{dm_{H_2}^e}{dt} = 4 \pi R_p^2 \phi_{\rm H_2} \mu_{\rm H_2}.
\label{eq:masslossrate}
\end{equation}
The loss rate depends on the time-dependent volume mixing flux of H$_2$ through $\phi_{\rm H_2}$.  In turn, the volume mixing ratio of H$_2$ is computed from the partial pressure of H$_2$ and the total surface pressure.  In the present work, the mass-loss rate for volatiles other than H$_2$ is set to zero.
\subsection{Mass exchange due to chemical reactions}
\label{app:chemreact}
During the magma ocean stage, the silicate melt and overlying atmosphere can accommodate mass exchange via chemical reactions.  We consider a general reaction with two reactants ($A$, $B$) and two products ($C$, $D$):

\begin{equation}
    aA + bB \rightleftharpoons cC + dD,
    \label{eq:reaction_general}
\end{equation}
where $a$, $b$, $c$, and $d$ are stoichiometric coefficients.  The reaction quotient $Q$ is a function of time $t$:

\begin{equation}
Q(t) = \frac{\{C\}^c \{D\}^d}{\{A\}^a \{B\}^b} \doteq \frac{Q_p(t)}{Q_r(t)},
\label{eq:equilibrium_constant}
\end{equation}
where curly brackets $\{ \}$ denote the activity of the species and the activity of a gas is its fugacity divided by a reference pressure.  At chemical equilibrium, the reaction quotient is equal to the equilibrium constant $K$, which is a function of temperature.  An appropriate temperature is that of the magma ocean--atmosphere interface; hence,

\begin{equation}
\frac{\{C\}^c \{D\}^d}{\{A\}^a \{B\}^b} - K(T) = 0.
\label{eq:Keqm}
\end{equation}
The reaction in Equation~\ref{eq:reaction_general} can also be expressed in terms of a mass balance of reactants and products:

\begin{equation}
m_A + m_B \rightleftharpoons m_C + m_D.
\label{eq:reaction_general_mass}
\end{equation}
This is because the fractional change of moles corresponds to the fractional change of mass, so correlating terms in Equation.~\ref{eq:reaction_general} and \ref{eq:reaction_general_mass}:

\begin{equation}
m_{B} = \left( \frac{b \mu_{B}}{a \mu_{A} } \right) m_A,\quad m_{C} = \left( \frac{c \mu_{C}}{a \mu_{A} }\right) m_A, \quad {\rm etc.}
\label{eq:reactionmasses}
\end{equation}
where $\mu$ is the molar mass.  It is convenient in Equation~\ref{eq:reactionmasses} to express the mass of the reactants and products using the mass of the first reactant ($m_A$) multiplied by a constant factor that depends only on molar masses and reaction stoichiometry; mathematically this choice is arbitrary, and another mass could equally be used.  Hence, each chemical reaction introduces a single unknown "reaction mass" ($m_A$) and one equation (Equation~\ref{eq:Keqm}) to the complete system of equations that we solve.  The functional dependencies of all quantities in Equation~\ref{eq:Keqm} are known, since we assume ideal gas behavior to connect gas activities to volatile partial pressures and the functional dependence of the equilibrium constant on temperature $T$ is documented in thermochemical tables \citep{JANAF}.

Evolution equations of Equations~\ref{eq:Keqm} and \ref{eq:reactionmasses} are necessary to time-step the model.  The time derivatives of the reactant and product masses are trivial to compute since stoichiometric coefficients and molar masses are constant.  For example (c.f., Equation~\ref{eq:reactionmasses}):

\begin{equation}
\frac{d m_B}{dt} = \left( \frac{b \mu_{B}}{a \mu_{A} } \right) \frac{d m_A}{d t},
\label{eq:dmBdt}
\end{equation}
and the time derivative of the reactant masses ($m_C$, $m_D$) can be similarly computed.  The time derivative of the equilibrium condition (Equation~\ref{eq:Keqm}) is:

\begin{equation}
    \frac{1}{Q_r} \frac{dQ_p}{dt} - \frac{Q_p}{Q_r^2} \frac{dQ_r}{dt} - \frac{dK}{dT} \frac{dT}{dt} = 0,
    \label{eq:dQdt}
\end{equation}
where the reaction quotient numerator $Q_p=\{C\}^c \{D\}^d$ and denominator $Q_r=\{A\}^a \{B\}^b$ depend on the volatile partial pressures and reaction stoichiometry.  The equilibrium constant $K$ depends only on temperature $T$ (Equations~\ref{eq:h_react} and \ref{eq:c_react}) and therefore $dK/dT$ is straightforward to evaluate.  The code also solves for $dT/dt$ in the mantle at each staggered mesh point, although only the change of surface temperature is required since this temperature defines the interface of the magma ocean and atmosphere.

Hence, to include reactions in the time stepper, Equation~\ref{eq:dQdt} constitutes an extra equation per reaction that must be satisfied, where the extra unknowns we solve for are the change of reaction mass for each reaction ($dm_A/dt$).  Equation~\ref{eq:dmBdt} relates the change of reaction mass for reactant $B$ and products $C$ and $D$ to the solution quantity ($dm_A/dt$).  For each volatile, we can thus evaluate the total rate of change of its mass due to all reactions that it participates in; this is the final term on the left-hand side of Equation~\ref{eq:volmassdt}.
\subsection{Initial volatile abundances}
\label{sect:initial_solver}
For the initial condition, we solve for the abundance of each volatile in the melt using a set of coupled equations consisting of all the volatile mass balances (Equation~\ref{eq:volmass}) and $r$ chemical reactions (Equation~\ref{eq:Keqm}).  To do this, estimates of the initial total volatile abundances $X_v^{\rm 0, est}$ are prescribed, and the initial atmospheric escape masses are set to zero ($m_v^e=0$).  We solve for the partial pressure of each volatile in the atmosphere and the reaction masses, the latter of which are likely to be nonzero.  Hence, it is convenient---although formally not necessary---to offset the estimates of the initial total volatile abundances ($X_v^{\rm 0, est}$) using the solved reaction masses to enforce that the initial total volatile abundances ($X_v^0$) adhere to chemical equilibrium constraints with zero reaction mass:

\begin{equation}
    X_v^0 = X_v^{\rm 0,est} - \frac{1}{M^m} \sum_w^r m_v^w.
\end{equation}
Hence, the reaction masses are zero by definition for the initial condition and can increase or decrease during the time integration.
\subsection{Alternative using thermodynamic components}
The system of volatiles we consider can be represented by two components: C and H$_2$, where we recall that oxygen is not explicitly tracked.  Hence, an alternative approach to solving the mass balance for each volatile is to instead solve for the total abundance of C and H$_2$ moles.  Conservation laws for C and H$_2$ abundances can be derived by converting the relevant volatile mass balances to C or H$_2$ moles and then summing.  Without chemical reactions, conserving species mass or abundance is the same because there is no exchange of material between species.  With reactions, either one can continue to track species mass by accounting for mass exchange by reactions (as we choose to do), or alternatively one can instead track the abundance of components.  Considering species mass allows one to retain the familiar mass formulation derived without reactions, but at the expense of more equations (e.g., five volatile species plus three reactions).  By contrast, considering components reduces the number of equations (e.g., two components plus three reactions), but requires a departure from the traditional species mass balance approach.  Ultimately, both approaches are valid and equivalent and will likely find utility in different applications.
\subsection{Simplified mass balance and interior depletion}
\label{app:simple}
The partitioning of volatiles between a fully molten magma ocean and atmosphere can be analyzed using a reduced form of the volatile mass balance (Equation~\ref{eq:volmass}).  This aids the interpretation of the results---notably demonstrating their universality---as well as providing analytical expressions and scaling relations.  We consider the mass balance for a single volatile that is only partitioned between the melt and atmosphere.  However, we retain the mean molar mass of the atmosphere $\bar{\mu}$ to provide insight when the atmosphere is dominated by the presence of another (or mixture) of volatiles that are less soluble than the volatile under consideration.  Following Equation~\ref{eq:volmass}:

\begin{equation}
X_v M^l + 4 \pi R_p^2 \left( \frac{\mu_v}{\bar{\mu}} \right) \frac{p_v}{|g|}  = X_v^0 M^m.
\label{eq:volmasssingle}
\end{equation}
For power-law solubility (Equation~\ref{eq:sol}),

\begin{equation}
\frac{\phi_g}{C} X_v + {X_v}^{\beta_v} = \frac{X_v^0}{C},
\label{eq:volmassnd}
\end{equation}
where $\phi_g = M^l/M^m$ is the global melt fraction and

\begin{equation}
C = \frac{4 \pi R_p^2 \mu_v {\alpha_v}^{-\beta_v}}{M^m \bar{\mu} |g|}.
\end{equation}
For a single species ($\bar{\mu}=\mu_v$), an Earth-sized mantle, and values of $\alpha_v$ and $\beta_v$ from the peridotite solubility law (Table~\ref{tab:h2o_sol}), the coefficient $C\approx 17/4$.
\subsubsection{Volatile depletion for $\beta_v=2$}
\label{app:beta2}
For the peridotite solubility law (as well as other laws), $\beta_v=2$ and Equation~\ref{eq:volmassnd} is a quadratic equation that can be solved for the volatile abundance in the melt:

\begin{equation}
X_v = \frac{-\phi_g}{2C} + \frac{1}{2} \sqrt{\frac{\phi_g^2}{C^2}+ \frac{4 X_v^0}{C}}.
\end{equation}
Hence, comparing to Equation~\ref{eq:volmassnd}, the scaled volatile mass in the liquid is

\begin{equation}
m_v^l =  \frac{-\phi_g^2}{2C^2} + \frac{\phi_g}{2C} \sqrt{\frac{\phi_g^2}{C^2}+ \frac{4 X_v^0}{C}},
\end{equation}
the scaled volatile mass in the atmosphere is

\begin{equation}
m_v^a =  \frac{X_v^0}{C} + \frac{\phi_g^2}{2C^2} - \frac{\phi_g}{2C} \sqrt{\frac{\phi_g^2}{C^2}+\frac{4 X_v^0}{C}},
\end{equation}
and the scaled total volatile mass is

\begin{equation}
m_v^t = \frac{X_v^0}{C}.
\end{equation}
Therefore, the volatile depletion $\mathcal{D}$ expressed as a fraction relative to the total volatile mass is

\begin{equation}
\mathcal{D} = 1 - \frac{m_v^l}{m_v^t} = \frac{m_v^a}{m_v^t} = 1 + \frac{\phi_g^2}{2CX_v^0} - \frac{\phi_g}{2X_v^0} \sqrt{\frac{\phi_g^2}{C^2}+\frac{4 X_v^0}{C}}.
\label{eq:depletion}
\end{equation}
Taking the derivative of Equation~\ref{eq:depletion} with respect to the initial volatile abundance $X_v^0$, we obtain

\begin{equation}
\frac{\partial \mathcal{D}}{\partial X_v^0} = \frac{\phi_g\left(-\phi_g \sqrt{\phi_g^2+4CX_v^0}+\phi_g^2+2CX_v^0\right)}{2(X_v^0)^2C\sqrt{\phi_g^2+4X_v^0C}}.
\end{equation}
All variables must be positive to be physically meaningful.  The denominator is clearly always positive and the numerator is also positive since the following condition is always satisfied:

\begin{equation}
\phi_g^2+2CX_v^0 > \phi_g \sqrt{\phi_g^2+4CX_v^0}\,.
\label{eq:depletion_cond}
\end{equation}
Therefore,

\begin{equation}
\frac{\mathcal{\partial D}}{\partial X_v^0} > 0\,.
\end{equation}
Hence, at a given melt fraction $\phi_g$, the relative depletion of the interior increases with the initial volatile abundance $X_v^0$ (Figure~\ref{fig:compare_depletion}).
\subsubsection{Volatile depletion for $\beta_v=1$}
For a volatile that obeys Henrian solubility behavior ($\beta_v=1$), similar to Appendix~\ref{app:beta2}, we can write:

\begin{equation}
X_v = \frac{X_v^0}{C+\phi_g}, \quad m_v^l= \frac{\phi_g X_v^0}{C(C+ \phi_g)}, \quad m_v^a = \frac{X_v^0}{C+\phi_g}.
\end{equation}
Hence, the interior depletion $\mathcal{D}$ is

\begin{equation}
\mathcal{D} = \frac{C}{C+\phi_g},\qquad \frac{\partial \mathcal{D}}{\partial X_v^0}=0.
\end{equation}
Notably, interior depletion is independent of the initial volatile abundance $X_v^0$ when $\beta_v=1$.
\section{Oxygen fugacity and atmosphere speciation}
\label{app:fO2}
\begin{figure*}[tbhp]
\centering
\includegraphics[height=0.9\textheight]{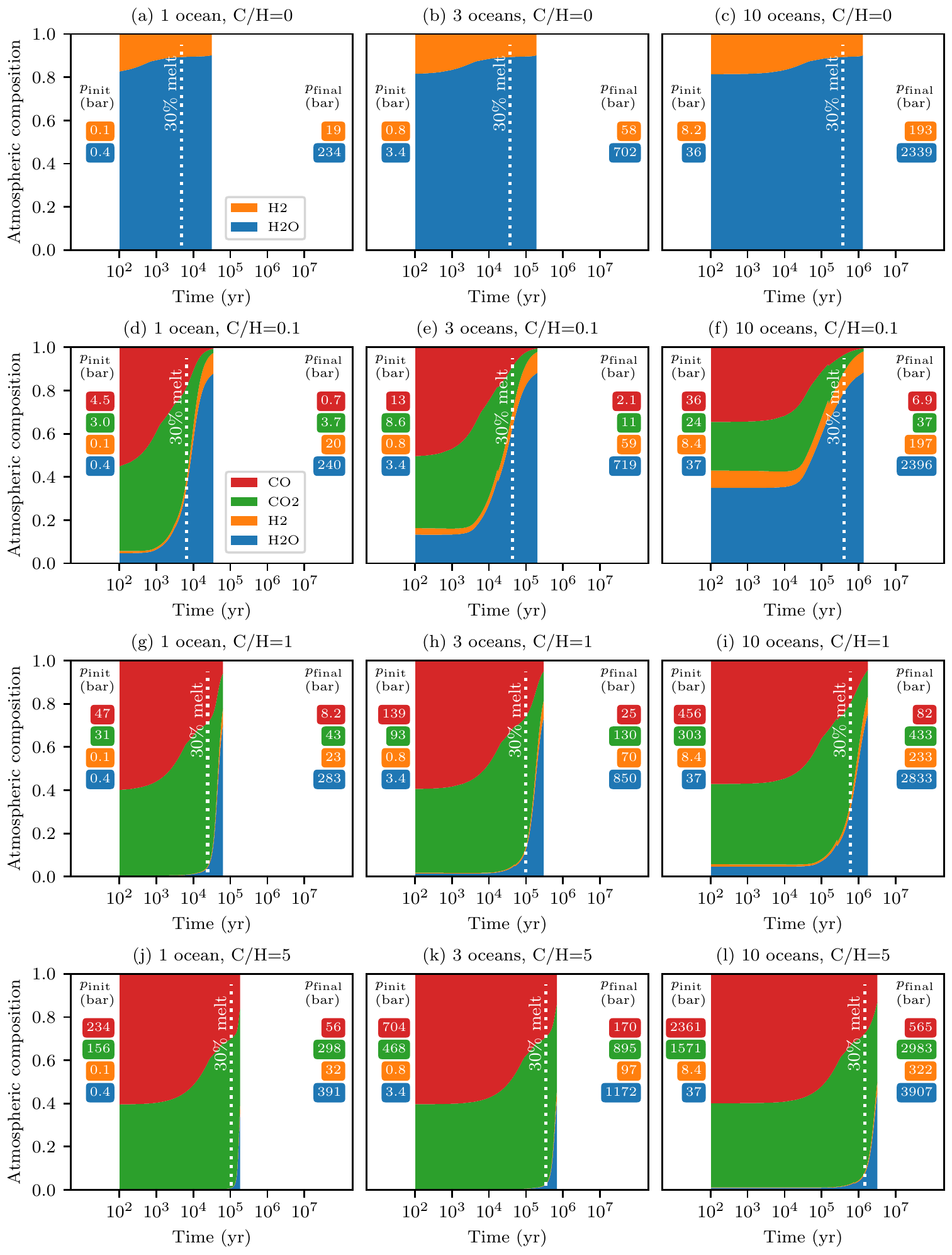}
\caption{Atmospheric composition during magma ocean outgassing at $f{\rm O_2}$ of $\Delta$IW=2.  See Figure~\ref{fig:ch_ratio} caption.}
\label{fig:ch_ratio_2fO2}
\end{figure*}

\begin{figure*}[tbhp]
\centering
\includegraphics[height=0.9\textheight]{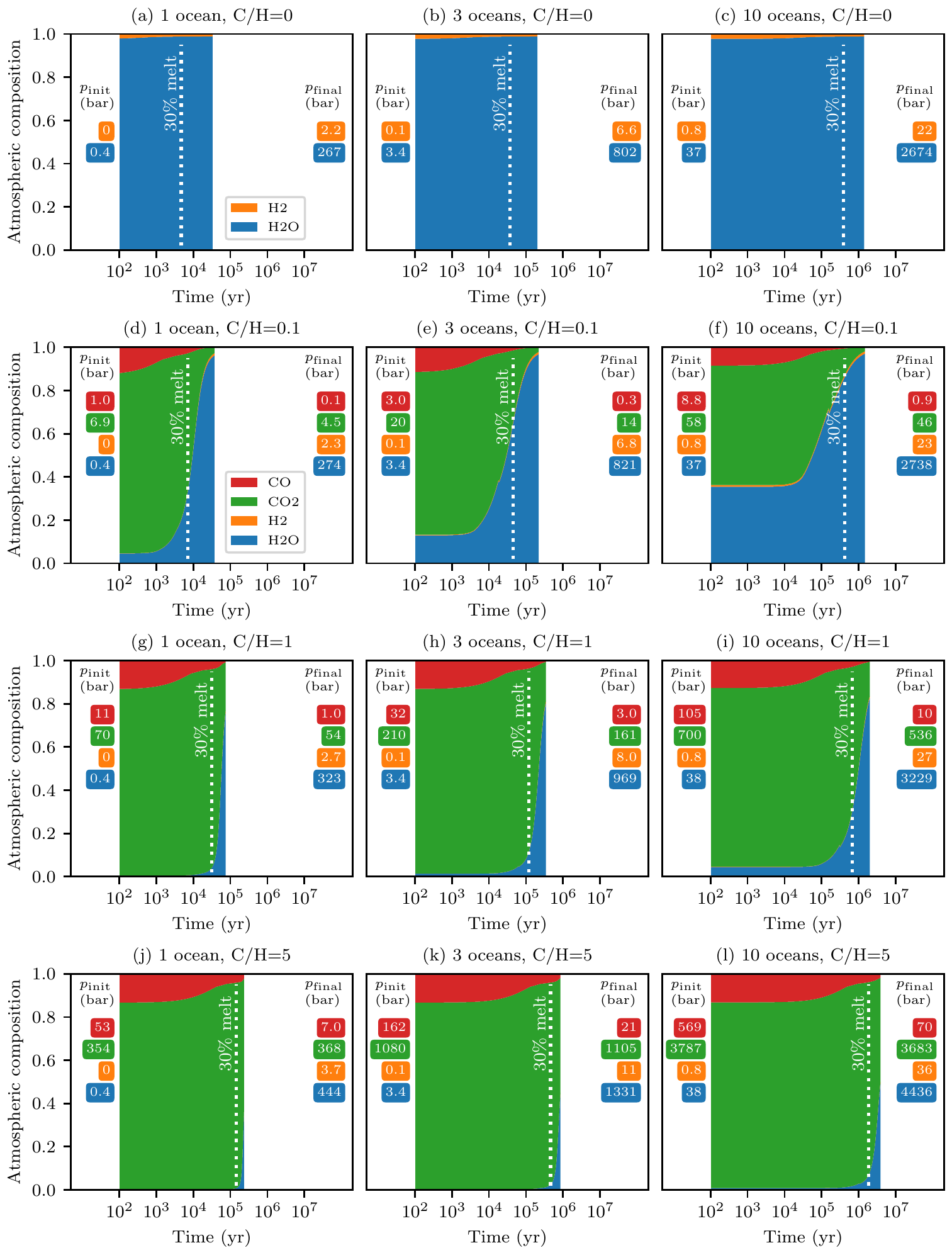}
\caption{Atmospheric composition during magma ocean outgassing at $f{\rm O_2}$ of $\Delta$IW=4.  See Figure~\ref{fig:ch_ratio} caption.}
\label{fig:ch_ratio_4fO2}
\end{figure*}

Figure~\ref{fig:ch_ratio} shows the atmospheric composition at an oxygen fugacity of $\Delta$IW=0.5.  The following figures show the same at $\Delta$IW=$-2$ (Figure~\ref{fig:ch_ratio_n2fO2}), $\Delta$IW=2 (Figure~\ref{fig:ch_ratio_2fO2}), and $\Delta$IW=4 (Figure~\ref{fig:ch_ratio_4fO2}).
\begin{figure*}[tbhp]
\centering
\includegraphics[height=0.9\textheight]{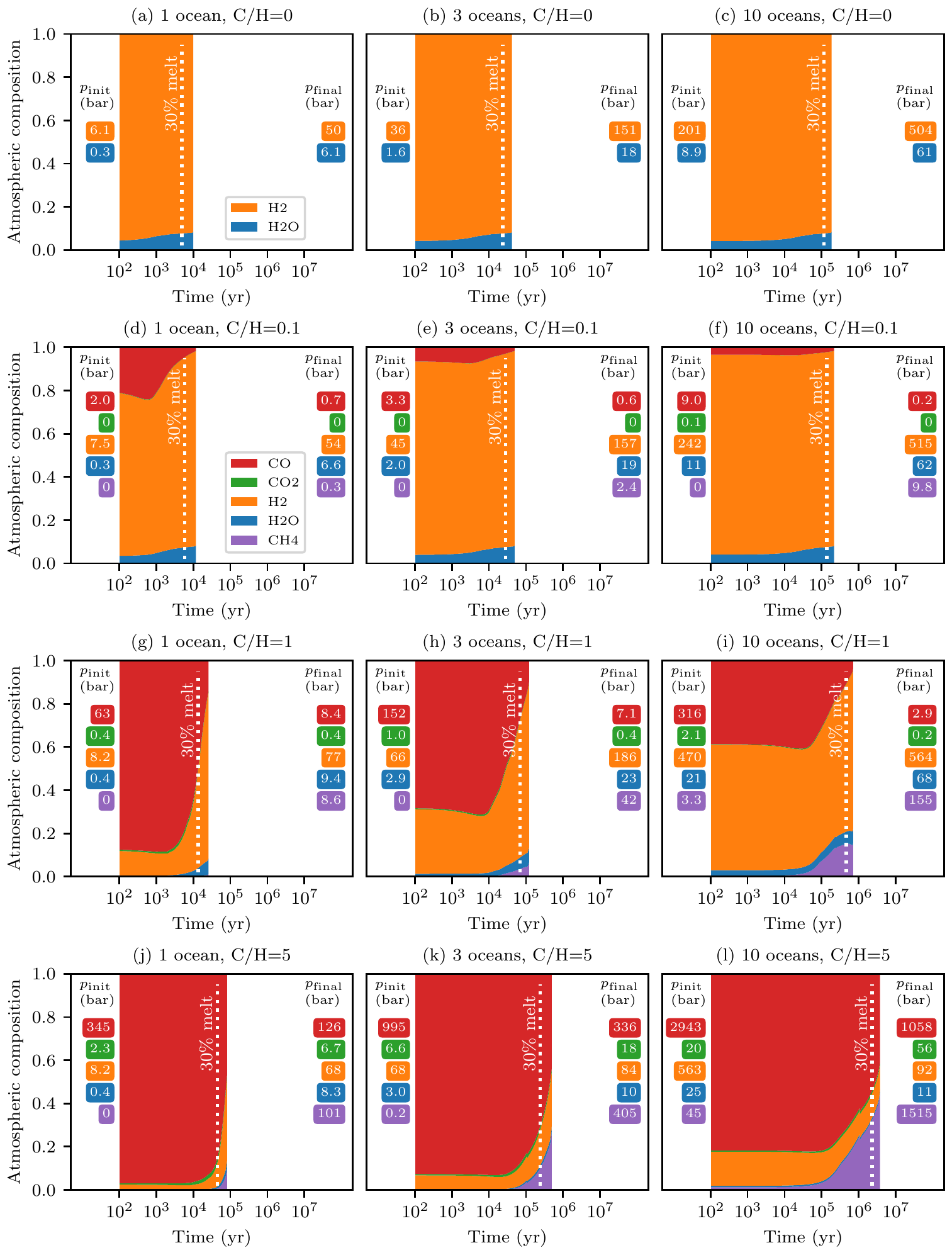}
\caption{Atmospheric composition during magma ocean outgassing at $f{\rm O_2}$ of $\Delta$IW=$-2$.  See Figure~\ref{fig:ch_ratio} caption.}
\label{fig:ch_ratio_n2fO2}
\end{figure*}
\section{Comparison with FactSage 8.0}
\label{app:factsage}
We independently verified the results of our model by comparison to FactSage calculations, assuming ideal gas behavior (Table~\ref{table:factsage}).

\begin{deluxetable*}{lll|lllll|lllll}
\centerwidetable
\tablecaption{Comparison of Select Outgassed Atmospheres Calculated Using Our Model Compared to FactSage Calculations\label{table:factsage}}
\tablewidth{0pt}
\tablehead{
\multicolumn{3}{c}{Case} & \multicolumn{5}{c}{Our model (basr)} & \multicolumn{5}{c}{FactSage (bars)}\\
\hline
\colhead{H Oceans} & \colhead{C/H} & \colhead{$f \rm O_2$ ($\Delta$IW)} & \colhead{CO} & \colhead{CO$_2$} & \colhead{H$_2$} & \colhead{H$_2$O} & \colhead{CH$_4$} & \colhead{CO} & \colhead{CO$_2$} & \colhead{H$_2$} & \colhead{H$_2$O} & \colhead{CH$_4$}}
\decimals
\tablecolumns{13}
\startdata
3 & 1 & -2.0 & 7.1 & 0.4 & 186 & 23 & 42 & 7.2 & 0.4 & 187.5 & 22.4 & 41\\
3 & 1 & +0.5 & 51 & 48 & 213 & 460 & 22 & 48.9 & 49.7 & 216.1 & 458.6 & 20.7\\
1 & 0.1 & +2.0 & 0.7 & 3.7 & 20 & 240 & 0 & 0.6 & 3.7 & 20.2 & 239.9 & 0.00043\\
1 & 5 & +4.0 & 7 & 368 & 3.7 & 444 & 0 & 6.4 & 367.6 & 3.7 & 445 & 0
\enddata
\tablecomments{This comparison reveals that the partial pressure of a given volatile differs by at most a few percent.}
\end{deluxetable*}

\clearpage





\end{document}